\documentclass[a4paper, fleqn,usenatbib]{mnras}
\usepackage{mathptmx}
\usepackage{mathptmx}
\usepackage{times, txfonts}
\usepackage[T1]{fontenc}
\usepackage{ae,aecompl}
\usepackage{graphicx}	% Including figure files
\usepackage{amssymb}	% Extra maths symbols
\usepackage[dutch, english]{babel}
\usepackage{mathrsfs}
\usepackage{latexsym}
\usepackage{longtable}
\usepackage{multicol}
\usepackage{supertabular}
\usepackage{multirow}
\usepackage{epsf}
\usepackage{color}
\usepackage{float}
\usepackage{enumerate}
\usepackage{natbib}
\usepackage{arydshln}
\usepackage{comment}
\usepackage{capt-of}

\newcommand{\todo}[1]{\textcolor{red}{[#1]}}

\newcommand{\ms}{m~s$^{-1}$}
\newcommand{\um}{$\mu$m}
\newcommand{\kgm}{kg~m$^{-3}$}

\newcommand{\mockalph}[1]{}

\title[]{The footprint of cometary dust analogs: \\ I. Laboratory experiments of low-velocity impacts and comparison with Rosetta data}
\author[L. E. Ellerbroek et al.]{
L. E. Ellerbroek$^{1}$\thanks{E-mail: ellerbroek@uva.nl},
B. Gundlach$^{2}$,
A. Landeck$^{2}$,
C. Dominik$^{1}$, 
J. Blum$^{2}$, 
\newauthor
S. Merouane$^{3}$,
M. Hilchenbach$^{3}$,
M. S. Bentley$^{4}$,
T. Mannel$^{4,5}$,
H. John$^{3}$,
\newauthor
H. A. van Veen$^{6}$
\\
$^{1}$ Astronomical Institute ``Anton Pannekoek'', University of Amsterdam, Science Park 904, 1098 XH Amsterdam, The Netherlands \\
$^{2}$ Institut f\"{u}r Geophysik und extraterrestrische Physik, Technische Universit\"{a}t Braunschweig,\\ Mendelssohnstrasse 3, D-38106 Braunschweig, Germany\\
$^{3}$ Max-Planck-Institut f\"{u}r Sonnensystemforschung, Justus-von-Liebig-Weg 3, D-37077 G\"{o}ttingen, Germany\\
$^{4}$ Space Research Institute, Austrian Academy of Sciences, Schmiedlstrasse 6, 8042 Graz, Austria\\
$^{5}$ Physics institute, University of Graz, Universit\"{a}tsplatz 5, 8010 Graz, Austria\\
$^{6}$ Electron Microscopy Center Amsterdam, Academic Medical Center, Amsterdam, The Netherlands 
}

\date{Accepted 2017 May 18. Received 2017 May 11; in original form 2017 March 27.}

\pubyear{2016}

\begin{document}
\label{firstpage}
\pagerange{\pageref{firstpage}--\pageref{lastpage}}
\maketitle

\begin{abstract}
Cometary dust provides a unique window on dust growth mechanisms during the onset of planet formation. Measurements by the Rosetta spacecraft show that the dust in the coma of comet 67P/Churyumov-Gerasimenko has a granular structure at size scales from sub-\um~up to several hundreds of \um, indicating hierarchical growth took place across these size scales. However, these dust particles may have been modified during their collection by the spacecraft instruments. Here we present the results of laboratory experiments that simulate the impact of dust on the collection surfaces of COSIMA and MIDAS, instruments onboard the Rosetta spacecraft. We map the size and structure of the footprints left by the dust particles as a function of their initial size (up to several hundred \um) and velocity (up to 6~\ms). We find that in most collisions, only part of the dust particle is left on the target; velocity is the main driver of the appearance of these deposits. A boundary between sticking/bouncing and fragmentation as an outcome of the particle-target collision is found at $v\sim2$~\ms. For velocities below this value, particles either stick and leave a single deposit on the target plate, or bounce, leaving a shallow footprint of monomers. At velocities $>2$~\ms and sizes $>80$~\um, particles fragment upon collision, transferring up to 50 per cent of their mass in a rubble-pile-like deposit on the target plate. The amount of mass transferred increases with the impact velocity. The morphologies of the deposits are qualitatively similar to those found by the COSIMA instrument.
\end{abstract}


%%%%%%%%%%%%%%%%%%%%%%%%%%%%%%%%%%%%%%%%%%%%%%%%%%

%%%%%%%%%%%%%%%%% BODY OF PAPER %%%%%%%%%%%%%%%%%%
\clearpage
\section{Introduction}

Comets are thought to be the most pristine bodies in the solar system surviving to this day. They are the remnants of the early stage of planet formation, during which planetesimals were formed. Being kilometre-sized bodies, they have overcome several growth barriers in a way that is as yet not fully understood by dust coagulation models \citep{Dominik2007, Johansen2014}. Therefore, studying the structure of the dust emitted from the surface of a comet is key to understanding these growth mechanisms. In particular, a key question is whether the cometary dust has a fractal structure, or is compact. A fractal structure would imply that cometary dust is a pristine remnant of hierarchical growth processes. A compact structure would indicate the dust has since its formation been modified by collisions or thermal processes.

The Rosetta mission to the comet 67P/Churyumov-Gerasimenko (hereafter 67P) has two instruments on board that are able to image individual dust particles smaller than 1 mm: the Cometary Secondary Ion Mass Anaylzer (COSIMA, \citealt{Kissel2007}) and the Micro-Imaging Dust Analysis System (MIDAS, \citealt{Riedler2007}). These instruments sample a complementary part of the parameter space. The optical microscopic imager COSISCOPE, part of the COSIMA instrument, has a spatial resolution of 14~\um~(enhanced to 10~\um~using sub-pixel sampling, \citealt{Langevin2016}). The MIDAS instrument is an atomic force microscope able to detect much smaller spatial scales (down to 4 nm in optimal configuration). MIDAS is limited in its ability to measure particles with heights greater than $\sim10$~\um~above the target plate.

%MIDAS
So far, the analysis of six particles detected by the MIDAS instrument have been published \citep{Bentley2016, Mannel2016}. These particles have sizes in the range 1~\um~up to a few tens of \um~(limited by the scan size, probably extending beyond that). The particles are aggregates of smaller grains with typical sizes between 0.5 and 3~\um. However, scans with the highest resolution analysed so far (80~nm) have revealed structures with ever smaller grain size down to 0.2~\um. Thus, it cannot be excluded that smaller substructures exist even below this size scale (Bentley et al., in prep.).

The six MIDAS particles show a hierarchical structure, with either compact or more porous packing of the grains. \citet{Mannel2016} present a fractal analysis of 2 different examples of these: a compact type (which, if it is fractal, would have a derived fractal dimension between 2 and 3), and a more porous type (with fractal dimension close to 1.7). These results suggests that the latter type of particle has formed through hierarchical growth of smaller grains, and has retained this property.

%COSIMA
The morphology of larger particles, as detected by COSIMA, can similarly be divided into two categories \citep{Merouane2016}. One category is referred to as `compact': single particles up to a few hundred \um~that appaear to not have fragmented or flattened on the target surface. A second category of particles are porous aggregates, which are divided into subtypes referred to as `rubble piles', `shattered clusters' and `glued clusters' by \citet{Langevin2016}. All of these subtypes are clusters of particles considered as part of a single collection event. Fragmentation of dust particles might also happen at an earlier stage, upon entry in the funnel. The latter is suggested by the detection of `showers' of particles: smaller particles confined to a small area of the instrument target, which were collected within a few days (the detector time resolution). These may be the result of a single breakup event of a larger, compact parent, that hit the funnel wall(s) upon entry. A key question is therefore whether the porous aggregates comprise an intrinsically separate dust population, or were in fact originally compact particles that fragmented upon impacting on the target at high velocity. 

%GIADA
A third instrument which studies the coma dust population of 67P is the Grain Impact Analyser and Dust Accumulator (GIADA, \citealt{Colangeli2007}), which is not equipped with an imager, but measures the momentum and cross-section of dust particles. This instrument thus provides constraints on the dust flux, velocity and density. Based on the dust collection by GIADA, \citet{DellaCorte2015} report the existence of two types of dust particles in the coma of 67P. On one hand there are detections of `compact' particles about 80-800~\um. These particles enter the instrument at velocities typically in the range 0.5~--~15~\ms, with a few detections up to 35~\ms~when the comet was near perihelion \citep{DellaCorte2016}. On the other hand, collections of intermittent showers of smaller particles are hypothesized to be the fragments of a low-density, fluffy component \citep{Fulle2015}. 

%Fulle2015: We associate the GDS fluffy fragments with the fluffy particles observed by COSIMA
%Mannel: These pebbles [pebble accretion pebbles] are linked with the com- pact agglomerates found by GIADA (Fulle et al. 2016b). [...] The smaller compact agglomerates detected by MIDAS like particle F might stem from the same pebble population, which would suggest a non-fractal structure. 
%Merouane: The classes defined as compact and porous aggregates can be slightly different for the two instruments and one cannot exclude that the COSIMA compact or aggregates are different from the GIADA compact and fluffy classifications, since both instruments measure different physical parameters. 

Discussion is ongoing about the association between the `compact' and `fluffy / porous' dust families detected by GIADA, COSIMA and MIDAS \citep{DellaCorte2015, Fulle2015, Merouane2016, Mannel2016}. However, no conclusive statement on this can be made, as these instruments measure different parameters of dust particles in different ranges via different collection methods. In reality, what is studied is the product of an interaction between the dust particles and the spacecraft or instrument. For instance, GIADA detects fluffy aggregates in showers, suggesting breakup of these particles shortly before entering the instrument \citep{Fulle2015}. Similarly, COSIMA seems to detect bursts, where within one day, one part of the cm-sized target is littered by many particles; this is best explained by the particles being in fact fragments of a larger particle that has broken up entering the instrument funnel. As noted by \citet{Merouane2016}, the COSIMA and GIADA `showers' are not coincident, suggesting they are both the result of particle fragmentation close to or within the instrument. For the MIDAS instrument, which also has a funnel entry, similar fragmentation is likely also happening although it is harder to confirm due to the limited spatial coverage used. 

This suggests a difference between the pre- and post-impact properties of the dust particles. A tool is therefore needed to simulate the entry of dust into the Rosetta instruments, and to study their impact products on the detector surfaces. By conducting experiments on targets used by both COSIMA and MIDAS, a direct comparison with Rosetta dust measurements is possible. Ultimately, the goal is to retrieve a full picture of the properties of the different dust populations of the coma of 67P, as detected by the instruments onboard the Rosetta spacecraft.

%FIGURE: SAMPLE
	\begin{figure}
\begin{center}
	\includegraphics[width=\columnwidth]{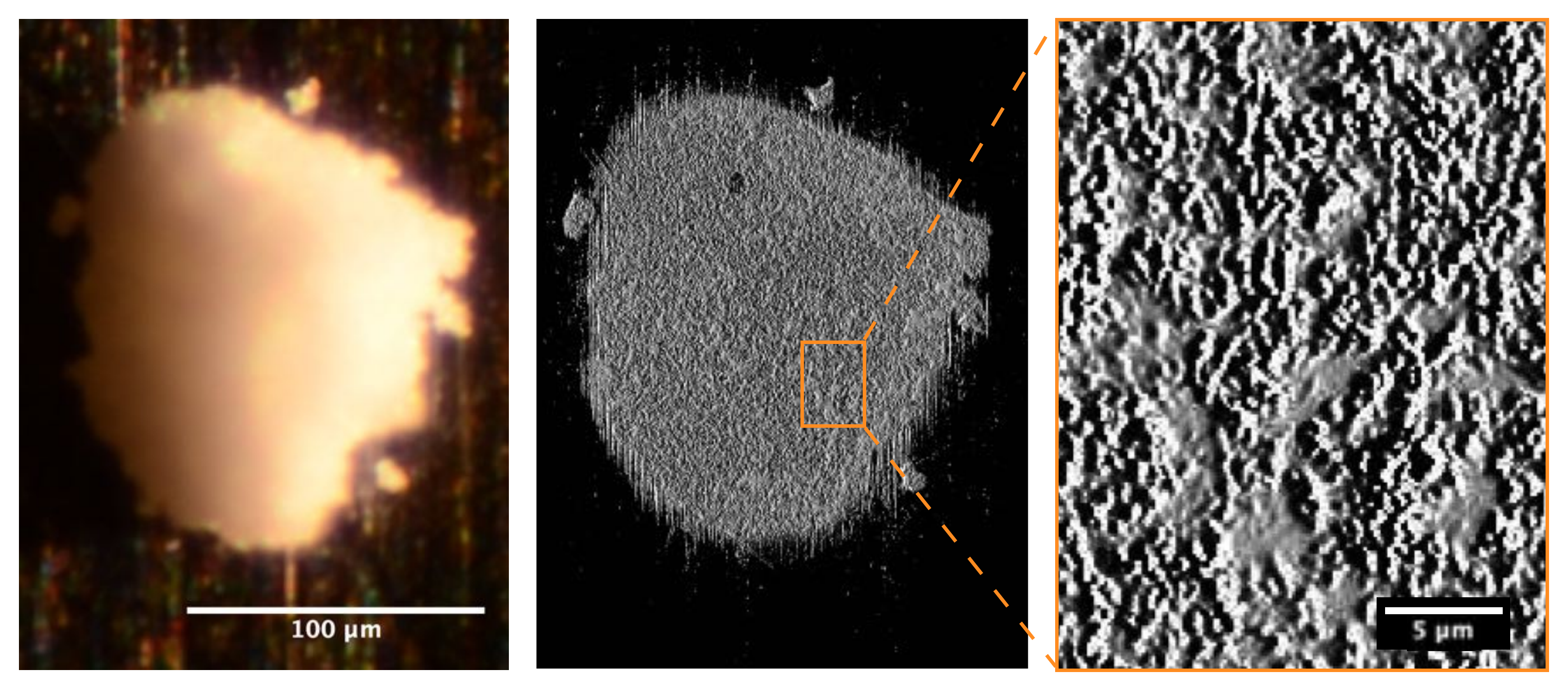}
	\includegraphics[width=.975\columnwidth]{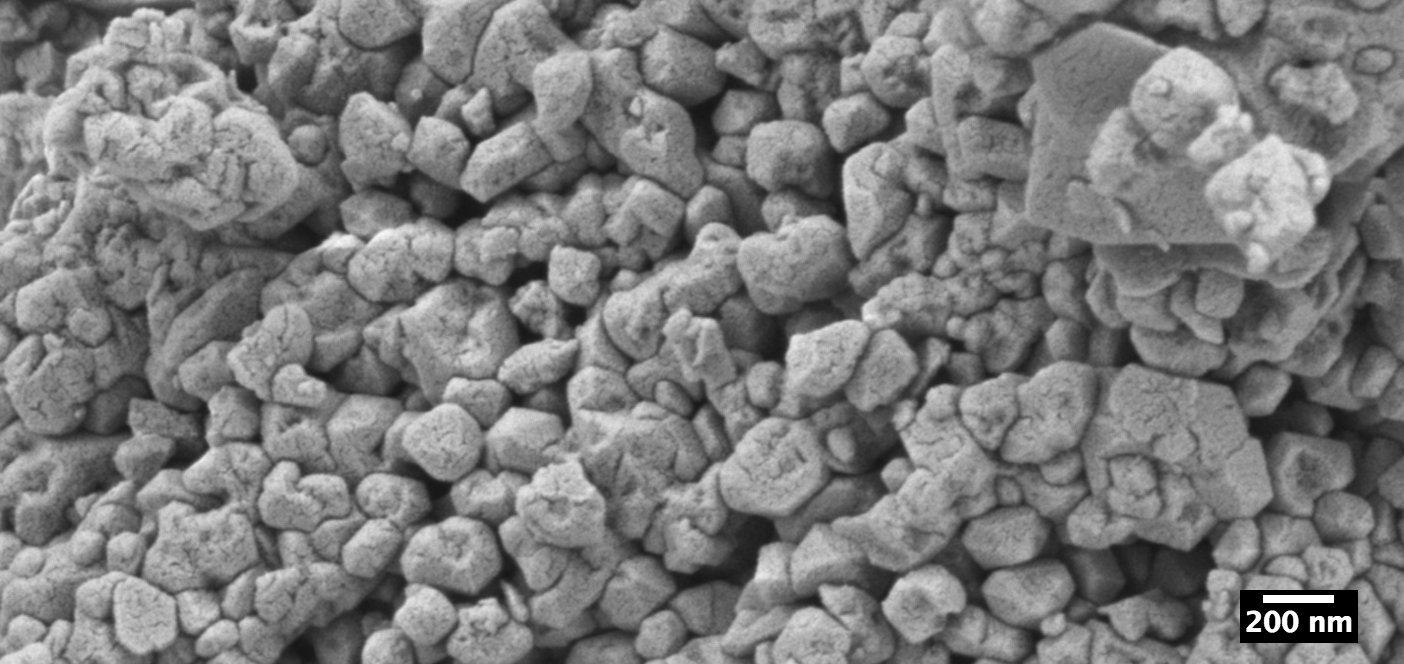}
\end{center}
    \caption{Images of an aggregate consisting of irregular polydisperse SiO$_2$. \textit{Top, from left to right:} optical microscope image (pixel size: 0.67~\um); 3D laser scanning confocal microscope image (pixel size: 0.14~\um); detail of middle panel. \textit{Bottom:} Scanning Electron Microscope image (dimensions: $4\times2$~\um) of a different aggregate of the same material.}
%Type name: Keyence VK-X200K 3D Laser Scanning Confocal Microscope  
    	\label{fig:sample}
	\end{figure}
	
In this paper, we present the results of an initial series of experiments that allow an interpretation of the results measured by the instruments onboard the Rosetta spacecraft. We provide a tool to link this multi-instrument data to the properties of the dust entering the spacecraft, by using an experimental setup in which dust aggregates collide on target surfaces. We use synthetic dust aggregates whose size, velocity and bulk density are comparable to the cometary dust aggregates measured by Rosetta. The collisions of the dust particles are monitored by a high-speed camera, and afterwards their morphologies are studied in detail by producing high-resolution images that can be compared with the Rosetta results. These `pre-impact' and `post-impact' analyses are compared to produce a link between what is seen by the Rosetta instruments, and the initial properties of the incoming dust. 
 
In Sect.~\ref{sec:methods}, a description of the test material, the experimental setup and the data analysis process are given. 
In Sect.~\ref{sec:results}, the results of our experiments are presented and compared with Rosetta data. 
In Sect.~\ref{sec:discussion}, we discuss the implications of our work with respect to the Rosetta results and look forward to further applications.
In Sect.~\ref{sec:conclusions}, we summarise our work and main results.

%FIGURE: SETUP PHOTOS
	\begin{figure*}
	\includegraphics[width=0.9\columnwidth]{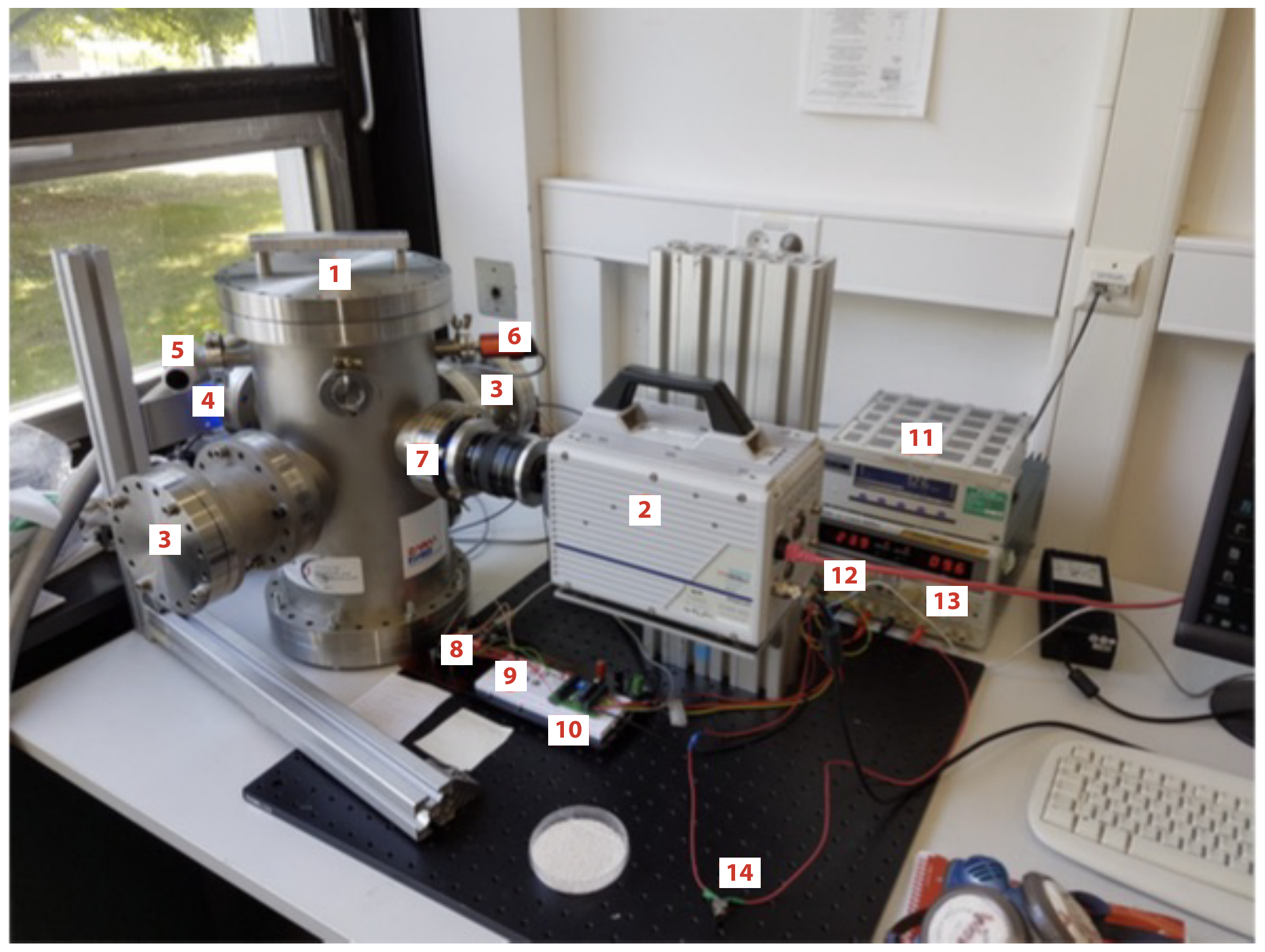}
	\hspace{0.5cm}
	\includegraphics[width=0.9\columnwidth]{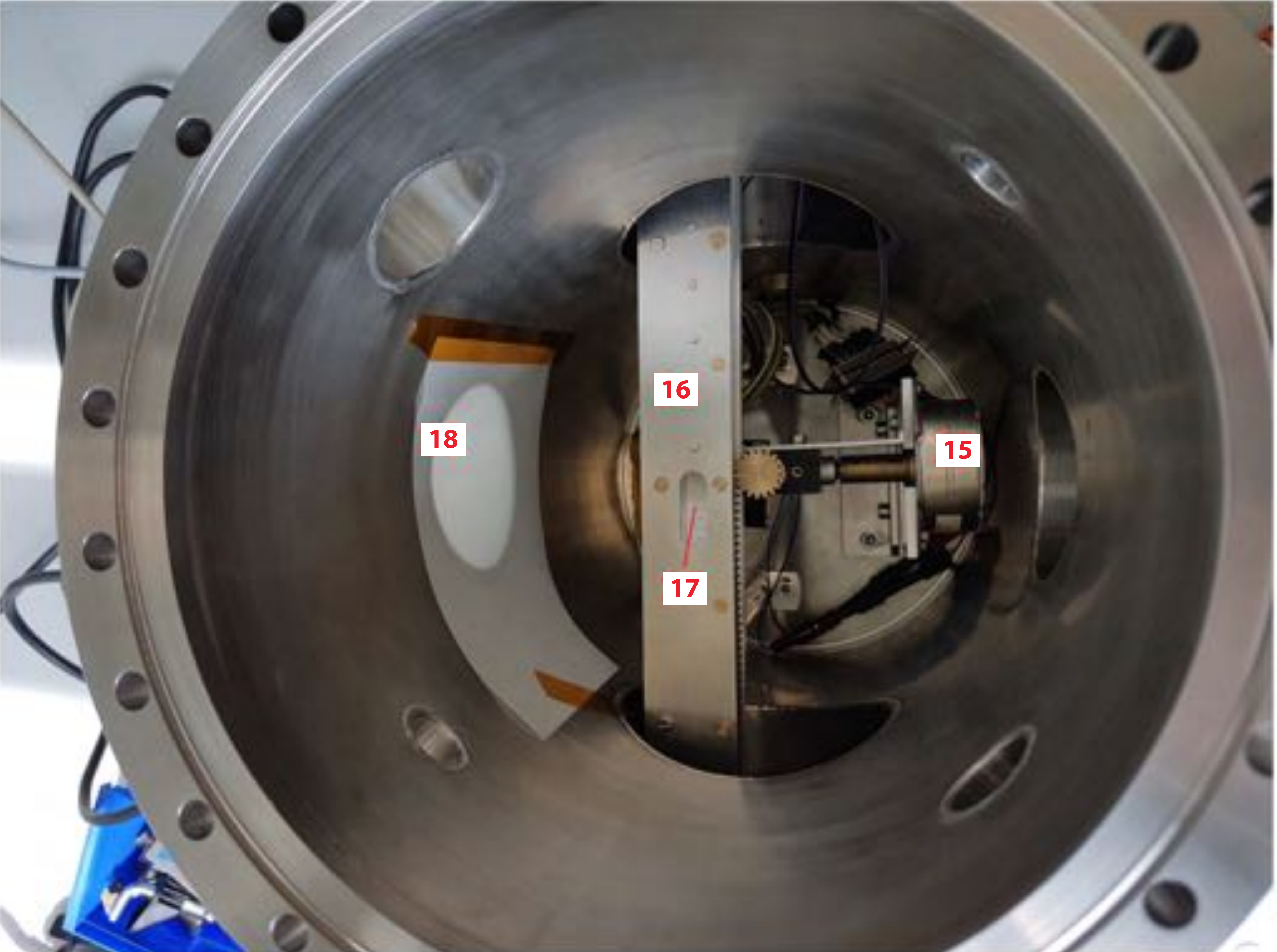}
    \caption{\textit{Left:} Photo of the experimental setup: 
    (1) Lid of vacuum chamber
    (2) High-velocity camera
    (3) Extensions containing dust cartridge
    (4) LED-array
    (5) Valve for pressurising the vacuum chamber
    (6) Pressure sensor
    (7) Window
    (8, 9, 10) Control panel for cartridge displacement
    (11) Pressure monitor
    (12, 13) Voltage meter for LED-array and lifting magnet
    (14) Release button for lifting magnet.
     \textit{Right:} Inside of the vacuum chamber (top view - in this image, dust particles launched would approach the observer):
     (15) Motor that controls displacement of dust cartridge
     (16) Container of dust cartridge
     (17) Opening through which dust particle is shot
     (18) Cover to achieve diffuse back-illumination by LED-array.
     }
    	\label{fig:setupphoto}
	\end{figure*}

%FIGURE: SETUP CARTOON
	\begin{figure}
	\includegraphics[width=\columnwidth]{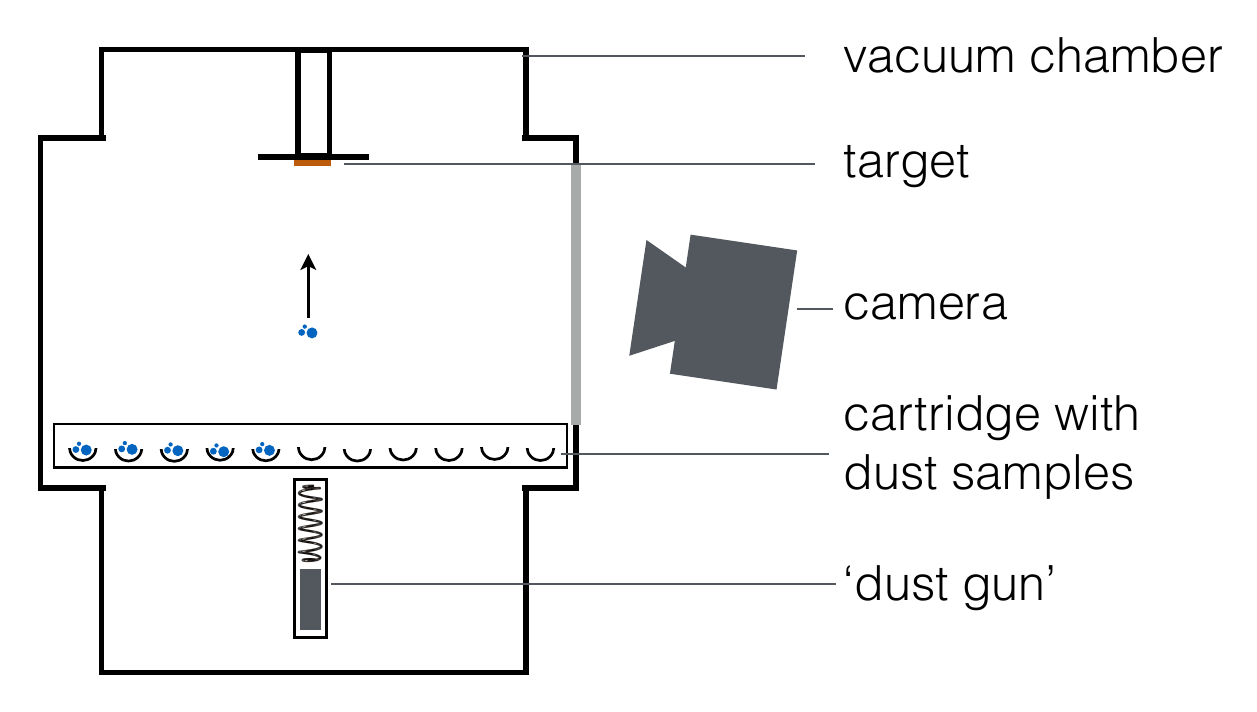}
    \caption{Schematic drawing of the inside of the vacuum chamber (side view).}
    	\label{fig:setupcartoon}
	\end{figure}

%SECTION: METHODS
\section{Methods}
\label{sec:methods}

In this section, we give a detailed description of our experiment. We describe the test material and compare its properties to cometary dust in Sect.~\ref{sec:methods:material}; we present the experimental setup in Sect.~\ref{sec:methods:experiment} and explain the data analysis process in Sect.~\ref{sec:methods:analysis}.

%SUBSECTION: MATERIAL AND PARAMETER RANGE
\subsection{Test material and parameter range}
\label{sec:methods:material}

For the series of experiments presented in this paper, we used a single type of synthetic dust aggregates. In this section we present the properties of these aggregates. Their degree of similarity with cometary dust is discussed in Sect.~\ref{sec:discussion:material}.

We used irregular-shaped, polydisperse SiO$_2$ with material density of $\rho_{\rm m}=2.6\times10^3$~\kgm (see Fig.~\ref{fig:sample}). The size range of the monomers is 0.1~--~10~\um, with the central 80 per cent of the mass in particles with sizes between 0.8 and 6~\um. The material was obtained from manufacturer Sigma-Aldrich; for a more detailed description see \citet{Blum2006, Weidling2012, Kothe2013}. As aggregates naturally are formed in the storage containers, we sieved the test material to obtain a particle diameter range of 100-400~\um. Since some aggregates fragmented on launch, particles $<100$~\um~were also included into the study.

The volume filling factor of the test material, after sieving, is $\phi=0.35\pm0.05$ \citep{Weidling2012}, resulting in a bulk density of  $\rho_{\rm b} \equiv \rho_{\rm m} \phi = (0.9 \pm 0.1) \times 10^{3}$~\kgm. Experimental studies \citep{Blum2004, Blum2006, Meisner2012, Lorek2016} show that the compressive strength of the material is a strong function of $\phi$, increasing from $10^4$~Pa up to $10^6$~Pa in the range $0.3<\phi<0.4$. Within the same range, the tensile strength is of order $10^3$~Pa, only slightly increasing as a function on $\phi$, while at higher values ($\phi>0.6$) the tensile strength may increase more steeply. The albedo of the material is close to 1, which affects the imaging of substructure with an optical microscope and grazing-angle illumination.
%The tensile strength is of order $10^3$~Pa as measured by \citet{Blum2006}.

%SUBSECTION: EXPERIMENTAL SETUP
\subsection{Experimental setup}
\label{sec:methods:experiment}

The setup of the experiment is shown in Figs.~\ref{fig:setupphoto} (photos) and \ref{fig:setupcartoon} (schematic drawing). The heart of the experiment is a vacuum chamber of $\sim 30$~cm in diameter. The vacuum chamber has two horizontal extensions, inside which a moveable dust cartridge is placed along a gear rack rail. The cartridge contains 20 holes of a few mm in diameter, in which a dust sample may be placed. Typically, a collection of $\sim 20$ particles was loaded into each hole. Below the hole is a piston, which launches the dust sample upwards when a current pulse is applied to a lifting magnet. The voltage of the magnet, cartridge position and launching are all controlled external to the vacuum chamber, so that multiple `shots' may be fired without having to pressurise the chamber in between. 

Upon loading the piston, a target plate is attached to an extension on the bottom of the lid, which places the target 8~cm directly above the lifting magnet and cartridge. Within the chamber a pressure of $\sim0.03 $~mbar is generated by a vacuum pump, in order to minimize the influence of the gas on the collisional outcome. The impact velocity of the dust particle is controlled by tuning the voltage of the lifting magnet, and measured afterwards on the camera images. The velocity is not seen to change significantly due to air drag or gravity over the particle's trajectory, as shown in Sect.~\ref{sec:analysis:preimpact}. %This is consistent with the stopping time due to air drag $\tau_{\rm stop}$=$\rho_{\rm s} a/\rho_{\rm air}v_{\rm th}$, and the free fall timescale $\tau_{\rm ff}$=$\sqrt{2d/g}$, exceeding the particle travel time from the piston to the target. % \todo{This is consistent with the stopping time using $\tau_{\rm stop}$=$\rho_{\rm s} a/\rho_{air}v_{\rm th}$, and the free fall time using $\tau_{\rm ff}$=$\sqrt{2d/g}$, and conclude that both are much larger than the travel time of the dust particle.}
%http://www.lpi.usra.edu/science/tom/aa_p1/node4.html

The particles are accelerated mechanically over a trajectory of $\Delta x\sim0.5$~cm up to a velocity $v$, after which the piston stops and the dust leaves the piston in the upward direction. To estimate whether the mechanical pressure is enough to break apart the particle during its launch, we compare the average pressure during acceleration to the tensile strength of a particle. For simplicity, we assume linear acceleration of a cubical particle with width $d$ and bulk density $\rho_{\rm b}$. The average pressure exerted on the contact surface $A$ is then
\begin{eqnarray}
%\langle P \rangle &=& \frac{m \langle a \rangle}{A} = \frac{d^3 \rho_{\rm b} \times \frac{1}{2} v^2 / \Delta x}{d^2}\\
%&=& 1~{\rm Pa} \times \frac{d}{100~\mu{\rm m}} \frac{\rho_{\rm b}}{10^3~{\rm kg~m}^{-3}}\left(\frac{\Delta x}{0.5~{\rm cm}}\right)^{-1}\left(\frac{v}{1~{\rm m~s}^{-2}}\right)^2
\langle P \rangle = 1~{\rm Pa} \times \frac{d}{100~\mu{\rm m}} \frac{\rho_{\rm b}}{10^3~{\rm kg~m}^{-3}}\left(\frac{\Delta x}{0.5~{\rm cm}}\right)^{-1}\left(\frac{v}{1~{\rm m~s}^{-2}}\right)^2
\end{eqnarray} 
At $v\sim10$~\ms, this value approaches the tensile strength of the polydisperse, irregular SiO$_2$ used in our experiments (see Sec.~\ref{sec:methods:material}). This is consistent with observations of particles breaking up upon launch at the high end of the velocity range ($v\sim5-6$~\ms). Their individual fragments are treated as separate particles in our analysis. The aggregates are not expected to compress due to the acceleration \citep{Blum2006, Guttler2009}.

A high-speed  video camera was used to record the dust impact on the target. Silhouettes of the dust particles were observed through the chamber window against a surface backlit by a LED array. Exposures of 0.05 ms were taken at a rate of 5000 frames per second. The focal depth was approximately 0.5 cm, and the camera was focused on the middle of the target plate. The spatial resolution of the images is 18.7~\um~per pixel. The camera was inclined at ab angle of $\sim5^\circ$, to be able to monitor both the trajectory of the dust particles, record the impact and identify the location of the deposit left by the particle on the target, as described in the next section. 

At the end of its trajectory, the particle hits the instrument target at an angle of incidence of typically $<5^\circ$ off-normal. The part that sticks to the target plate, including scattered loose components that clearly originate from the same particle, is referred to as a `deposit'. The analysis of these deposits is described in the next subsection. For the experiments presented in this paper we used the 0.5~mm thick and 1$\times$1~cm wide silver plates without coating, which are also used in the COSIMA instrument \citep[`silver blank',][]{Langevin2016}. Tests were also conducted with `black gold' (COSIMA) and solgel (MIDAS) targets, with similar results that justify the usage of silver plates (see Sect.~\ref{sec:discussion:comparison} and \citealt{Hornung2014}).

% SUBSECTION: DATA ANALYSIS
\subsection{Data analysis}
\label{sec:methods:analysis}

This section describes the data analysis phase. The analysis consists of two parts: `pre-impact' (retrieval of dust particle size and velocity) and `post-impact' (retrieval of deposit size and morphology). A consistent method was used to match the particles identified in the video to the deposits on the target.

%FIGURE: PRE-IMPACT TRACING AND COLLISIONS
%	\begin{figure}
%	\includegraphics[width=\columnwidth]{examplemovie.pdf}
%	\includegraphics[width=\columnwidth]{S11_48frame.jpg}
%	\includegraphics[width=\columnwidth]{fig_impactor.pdf}
%    \caption{Pre-impact analysis. \textit{(a)} Stills from a sample video; time since launch indicated in top left corner 
%		\textit{(b)} Still of a video (different experiment than above), with lines indicating multiple traced particles, whose size and velocity were %determined. Grayscale is inverted as compared to top and bottom panels.
%    		\textit{(c)} Stack image of 40 frames (8 ms) of the collision of a particle ($d_{\rm pre} \sim 360\pm20$~\um, $v$=$2.6\pm0.6$\ms) that %approaches the target from below and fragments upon impact. }
%    	\label{fig:preimpact}
%	\end{figure}
%

%FIGURE: PRE-IMPACT TRACING AND COLLISIONS
	\begin{figure*}
	\includegraphics[width=.75\textwidth]{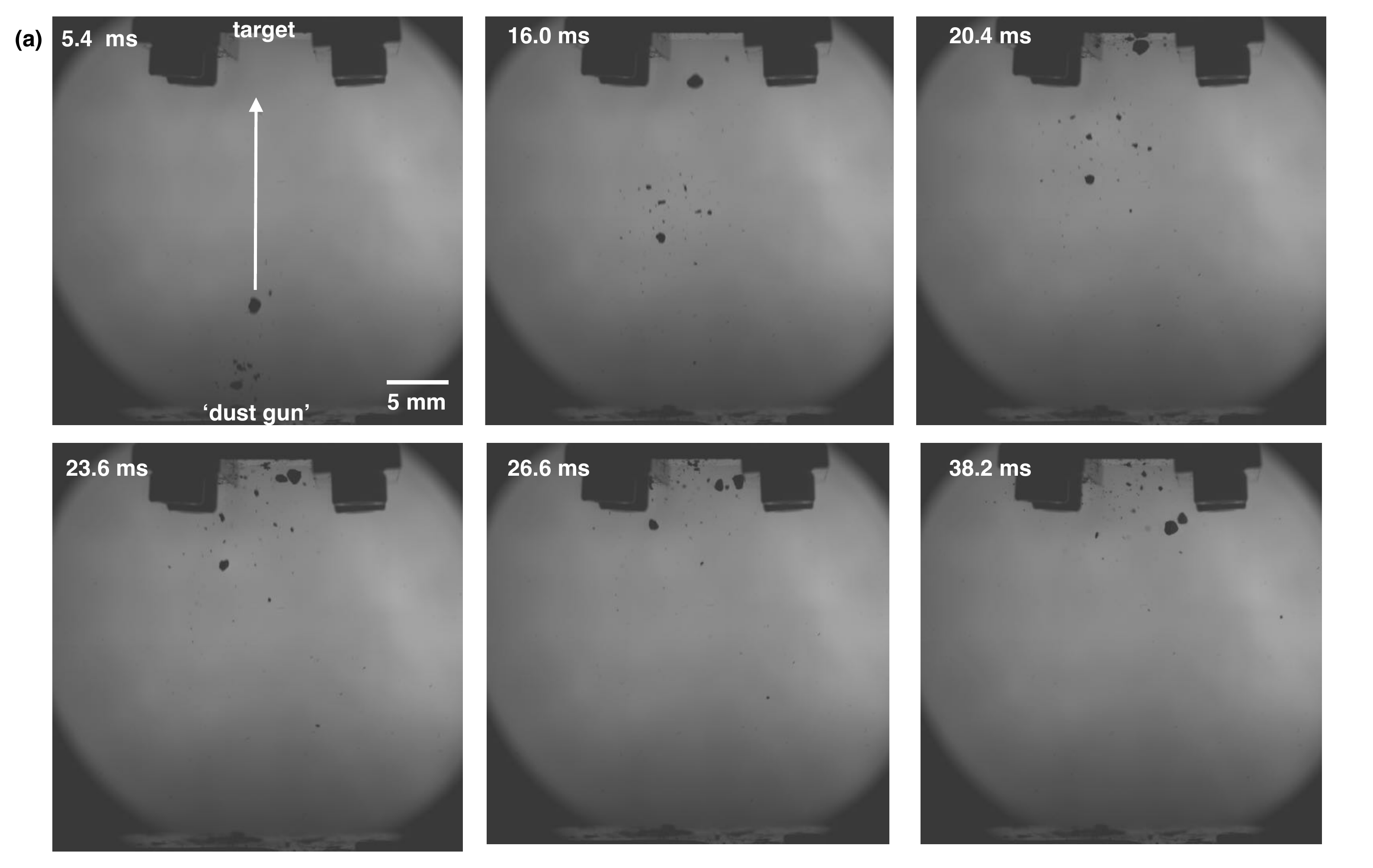}\\
	\includegraphics[width=\columnwidth]{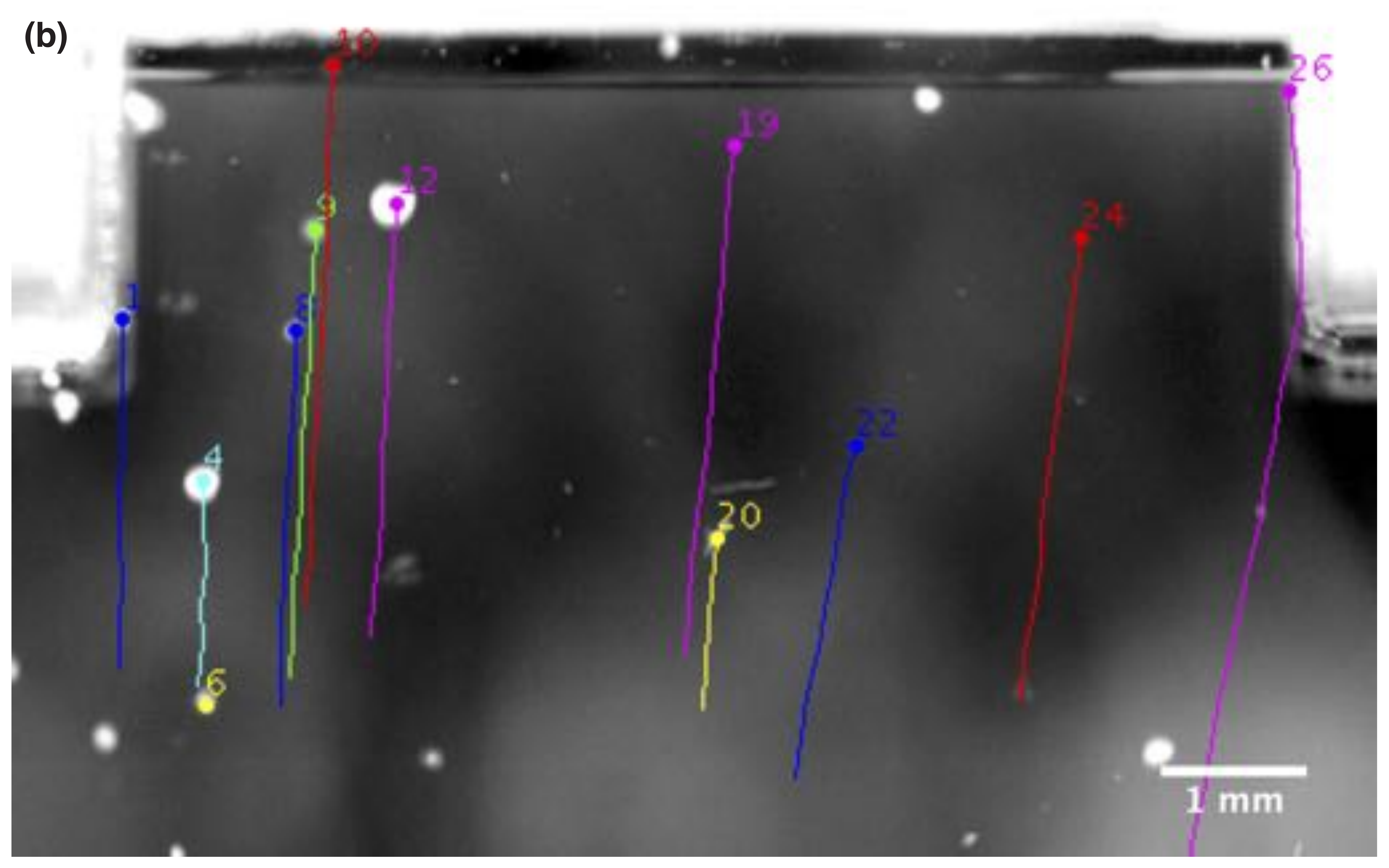}
	\includegraphics[width=\columnwidth]{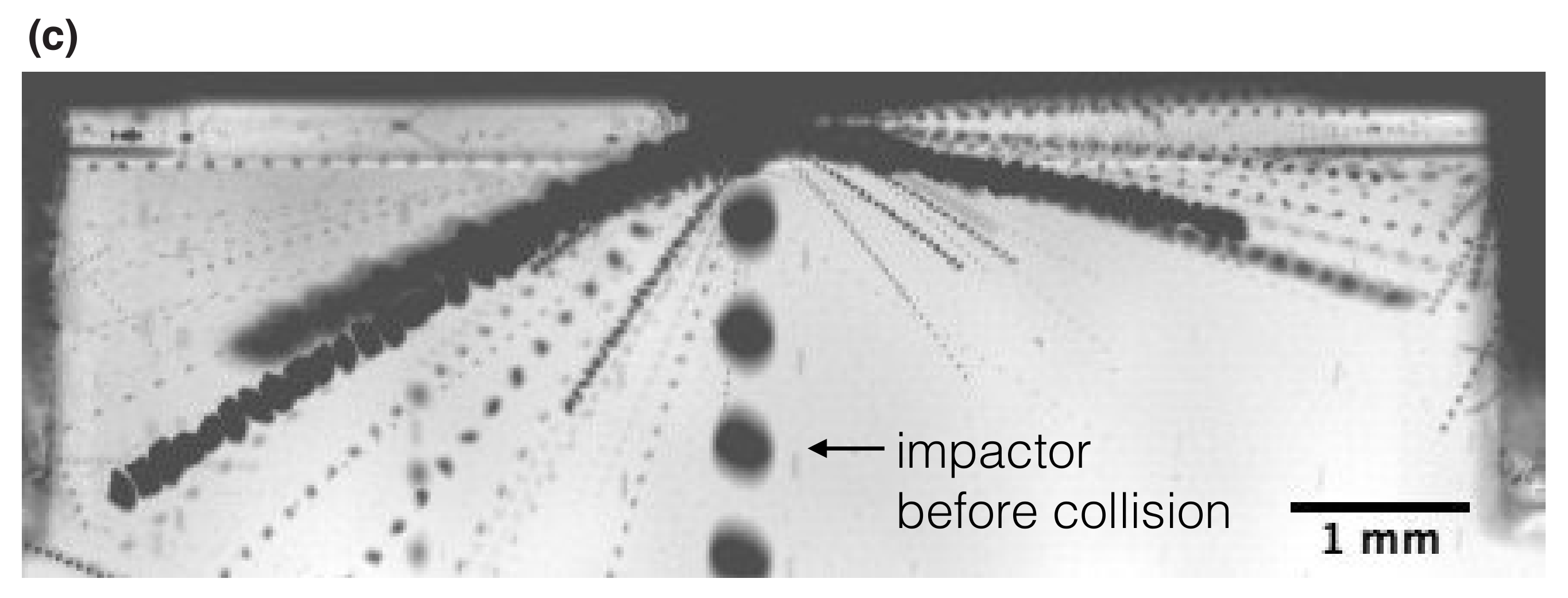}
    \caption{Pre-impact analysis. \textit{(a)} Stills from a sample video; time since launch indicated in top left corner 
		\textit{(b)} Still of a video, with lines indicating multiple traced particles, whose size and velocity were determined. The brightness scale has been inverted for better visibility.
    		\textit{(c)} Stack image of 40 frames (8 ms) of the collision of a particle ($d_{\rm pre} \sim 360\pm20$~\um, $v$=$2.6\pm0.6$\ms) that approaches the target from below and fragments upon impact. Note that the images in these three panels are from different experiments.}
    	\label{fig:preimpact}
	\end{figure*}

% TABLE: EXPERIMENTS
\begin{table}
	\centering
	\caption{Summary of pre-impact analysis. For targets S2 and S5, two shots were fired and analysed.}
	\label{tab:experiments}
	\begin{tabular}{cccc} % four columns, alignment for each
		\hline
		Target & \#paricles & Size range ($\mu$m) & Velocity range (m s$^{-1}$) \\
\hline
S1	& 11 & $35-340$ & $0.4-1.5$ \\
S2	& 44 (22 + 22) & $35 - 340$ & $0.4 - 1.7$  \\
S3  & 6 & $130-410$ & $2.6-3.4$ \\
S4 	& 8 & $35-130$ & $0.3-3.8$ \\
S5	& 17 (7 + 10) & $35-360$ & $2.1-3.9$ \\
S6	& 1 & $370$ & $4.9$  \\
S7 	& 23 & $30-220$ & $4.3-6.0$ \\
\hline
Total & 110 & $30-410$ & $0.3-6.0$ \\
\hline
	\end{tabular}
\end{table}

\subsubsection{Pre-impact analysis}
\label{sec:analysis:preimpact}

Videos of 9 shots on 7 different targets were analysed. As multiple aggregates were loaded upon each piston, and (at high velocities) some fragmented upon launch, after a single shot typically 40-50 particles in the sizes up to 400~\um~are visible on the video images (see Fig.~\ref{fig:preimpact}a). In total, we identified and analyzed 110 unique particles on these videos (see Tab.~\ref{tab:experiments}). 

Two parameters are derived from the videos: particle impact velocity and particle size (see \ref{tab:experiments}). To retrieve impact velocity $v$, the individual particles that impact on the plate were tracked (see Fig.~\ref{fig:preimpact}b). The particles were identified on the last 15 to 20 frames before impact. Over this trajectory ($\sim 1$ cm), the velocity decrease due to gravity and air pressure is seen to be on the order of 0.1 \ms ($1-10$ per cent of the absolute velocity). Thus, the average velocity over the last 15 to 20 frames yield a satisfactory estimate of the impact velocity. 

If a particle fragments upon impact, the fragments are seen to scatter on approximately straight trajectories for the next few microseconds, see Fig.~\ref{fig:preimpact}c. This indicates that during impact, the collision dynamics are not significantly altered by gravity. Furthermore, once part of the particle sticks to the target, this deposit retains its morphology despite it facing downwards. This shows that the contact force between particle and target far exceeds gravity. These effects demonstrate that, for the purpose of studying the collision dynamics and sticking of dust aggregates, our experimental setup gives a good approximation of the circumstances in space.

%FIGURE: PRE- AND POST-IMPACT SIZE MEASUREMENT 
	\begin{figure}
\begin{center}
		\includegraphics[width=0.85\columnwidth]{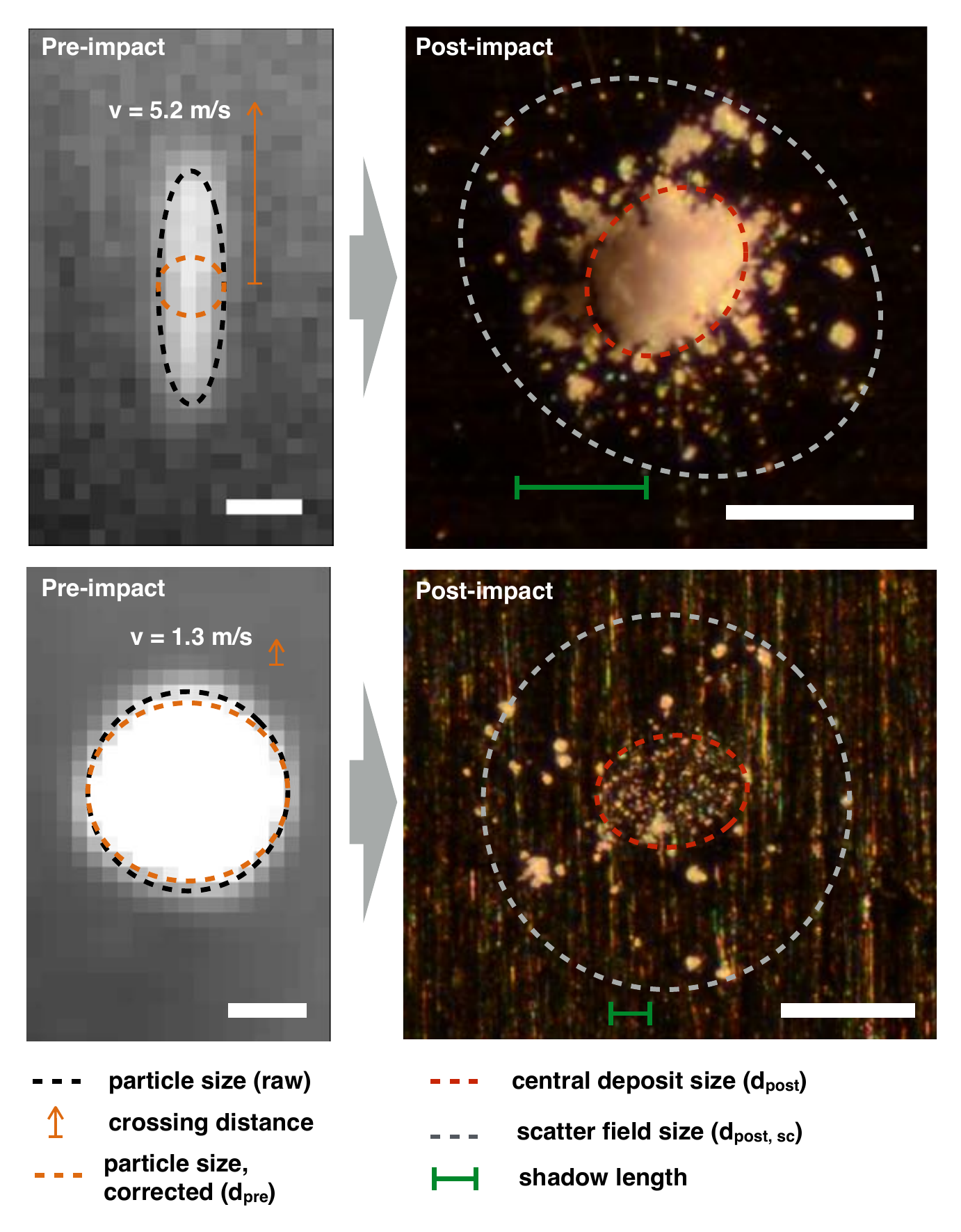}
\end{center}
    \caption{Size measurements made pre-impact (left) and post-impact (right) on a fast-moving (top) and slow-moving (bottom) particle. Due to their movement, the particles are smeared out in the vertical direction; this is corrected by the method described in Sect.~\ref{sec:analysis:preimpact}. The scale bar length in each image is 100~\um. The brightness scale of the images on the left has been inverted for better visibility. The shadow of the deposits on the images on the right was measured on a separate image with a modified brightness scale, similar to the method described in \citet{Langevin2016}.}
    	\label{fig:sizemeasurement}
	\end{figure}
	
	%FIGURE: TARGET
	\begin{figure*}
	\includegraphics[width=0.8\textwidth]{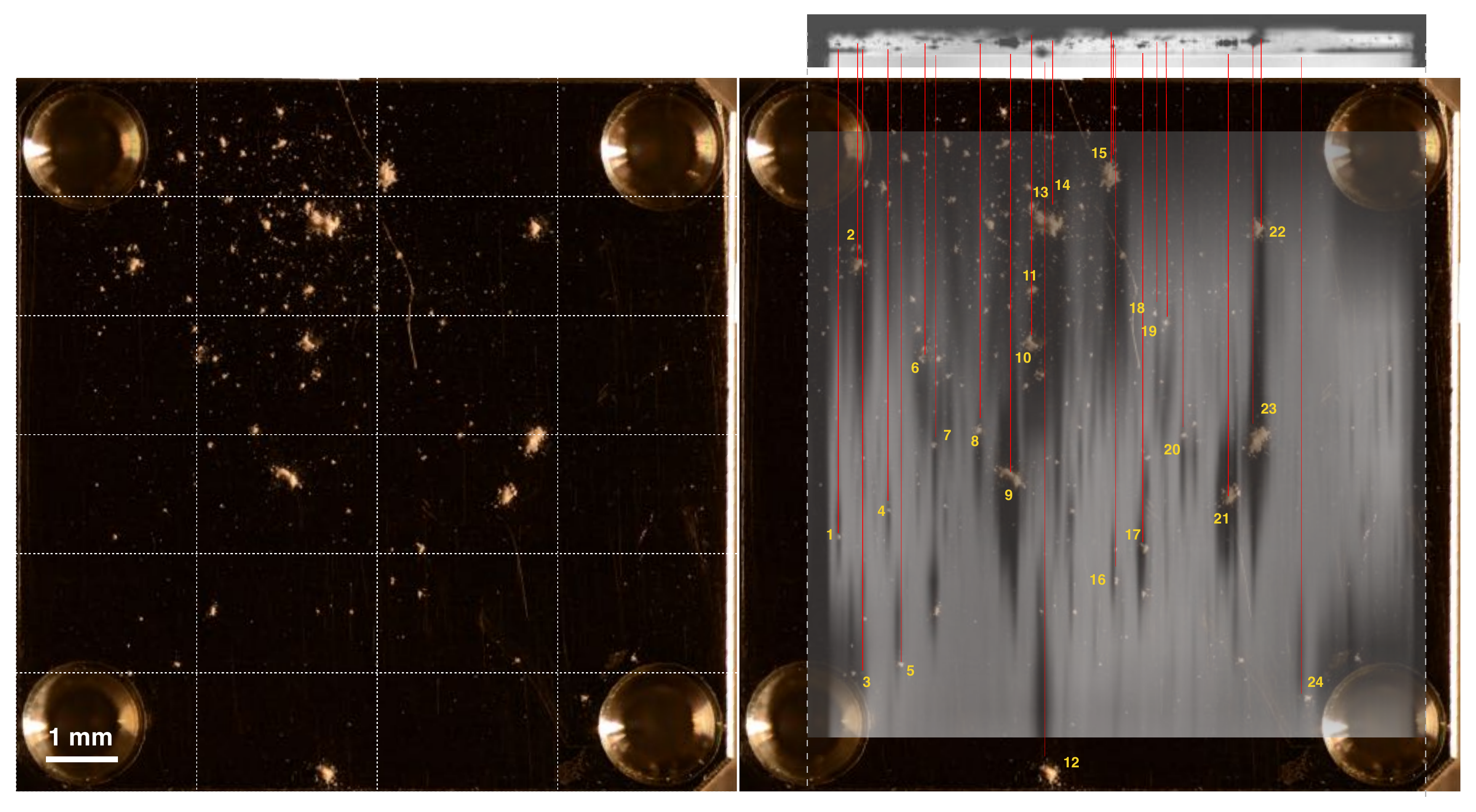}
    \caption{\textit{Left:} Microscope image of a target (S7, $v$ = $4.3 - 6.0$~m~s$^{-1}$) post-impact. The gridlines indicate the images from which the mosaic was compiled. \textit{Right:} The last frame of the corresponding experiment video (\textit{top right}) stretched and overlaid on the microscope image. Lines are drawn to illustrate the match between particles and their deposits. Only particles with a clear counterpart in the microscope image were matched and considered in the analysis.}
    	\label{fig:target}
	\end{figure*}
	
The size of the dust particles was measured by drawing an ellipse around the silhouette of the particle on the image (Fig.~\ref{fig:sizemeasurement}). While the morphology of the particles was irregular, the shapes approximated spherical reasonably well. We measure the ellipse that encompasses the particle on $\sim 5$ different frames preceding impact. At high velocities, the images are significantly blurred in the vertical direction. For example, at $v=5$~\ms, within the 0.05 ms exposure time the crossing distance of a particle is $\sim140$~\um. We correct for the blurring subtracting this amount from the size measurement in the vertical direction. We summarise the pre-impact size of the particles in the parameter $d_{\rm pre}$, being the average of these ellipse axes. Taking the image spatial resolution and the uncertainties introduced by lighting and focal depth into account, we adopt an error of 20 \um~for $d_{\rm pre}$. The relative measurement error in $d_{\rm pre}$ is thus close to unity for particles smaller than 50~\um, $\sim20$~per cent for 100~\um-particles, and $<15$~per cent for particles larger than 150~\um. 

An estimate of the pre-impact mass, $m_{\rm pre}$, of the particle is obtained by approximating it as a sphere with diameter $d_{\rm pre}$.The smallest particle size we are able to investigate is $\sim30$~\um, which is limited by the video resolution.  For very small particles ($d_{\rm pre} \lesssim 50$~\um) that have velocities >4~\ms and are thus significantly blurred, the value of $m_{\rm pre}$ should be considered an upper limit. 

%\begin{equation}
%m_{\rm pre} = \frac{\pi}{6} \rho_{\rm b} d_{\rm pre}^3
%\label{eq:mpre}
%\end{equation}
%with $\rho_{\rm b}$ the equivalent density as defined in Sect.~\ref{sec:methods:material}. 

%FIGURE: S17 ZPROJECT

\subsubsection{Post-impact analysis and matching}
\label{sec:analysis:postimpact}

Immediately after a successful experiment, the chamber was slowly pressurised and the target imaged in an optical microscope with a 5x magnifying lens at a resolution of 0.61~\um~per pixel. An example of a post-impact target is given in Fig.~\ref{fig:target}. On these mosaic images, multiple deposits were identified and labelled. A LED light was placed at a distance of 9 cm at an angle of 16$^\circ$; measuring the shadow cast by the dust particles allowed for an estimate of the height of the particle to be made. The pre-impact and post-impact dust particles were matched by rectifying and overlaying the final frame of the video image over the post-impact target image. Only particles that have a clear match with a measurable counterpart in both video and microscope images were included into this analysis. This does not introduce a selection effect, as non-detection is random and does not result in the exclusion of particles. No cases were observed where a clearly visible collision did not lead to a deposit. %, with the exception of a few impacts at very low angle of incidence during test measurements.

Three different post-impact dimensions were measured on the microscope images (see Fig.~\ref{fig:sizemeasurement}, right). An ellipse was drawn manually containing the largest central fragment of the deposit (i.e. the central part covering the target, or in the case of shallow footprints, where the particle density is highest). The average of the major and minor axes of this ellipse is defined as the deposit diameter ($d_{\rm post}$). A second, wider, ellipse was drawn around the deposit containing all scattered fragments disconnected from the central fragment but clearly related to the same collision event. The average of the major and minor axes of this ellipse is defined as the diameter of the scatter field ($d_{\rm post, sc}$). Finally, the length of the shadow cast by the central deposit was measured and deprojected to produce the maximum deposit height ($h_{\rm post}$). The shadow length was measured from the location of the highest point of the central fragment, which was in most cases easily identifiable from the microscope images by modifying the brightness scale. The post-impact dimensions $d_{\rm post}$, $d_{\rm post, sc}$ and $h_{\rm post}$ are measured with an uncertainty of $\sim 10$~per cent for small particles ($\sim 50$~\um) decreasing to a few percent for large particles ($> 150$~\um). 

These dimensions result in an estimate of the deposit mass:
\begin{equation}
m_{\rm post} = \frac{\pi}{4} \epsilon \rho_{\rm b} d_{\rm post}^2 h_{\rm post}
\end{equation}
where $\epsilon$ is the volume filling factor of the cylinder encompassing the central fragment. By simple geometry, we adopt $\epsilon=0.33$ for a pyramid-shaped, $\epsilon=0.67$ for hemispherical-shaped, and $\epsilon=1$ for a flat, shallow central fragment. We neglect the contribution by scattered fragments outside of $d_{\rm post}$, and assumed that no compression of the material occurs (which would enhance the equivalent density). Both of these effects contribute to $m_{\rm post}$ being an underestimate of the true deposit mass. 

%FIGURE: PARTICLEPLOTS MORPHOLOGY
	\begin{figure*}
	\includegraphics[height=0.8\columnwidth]{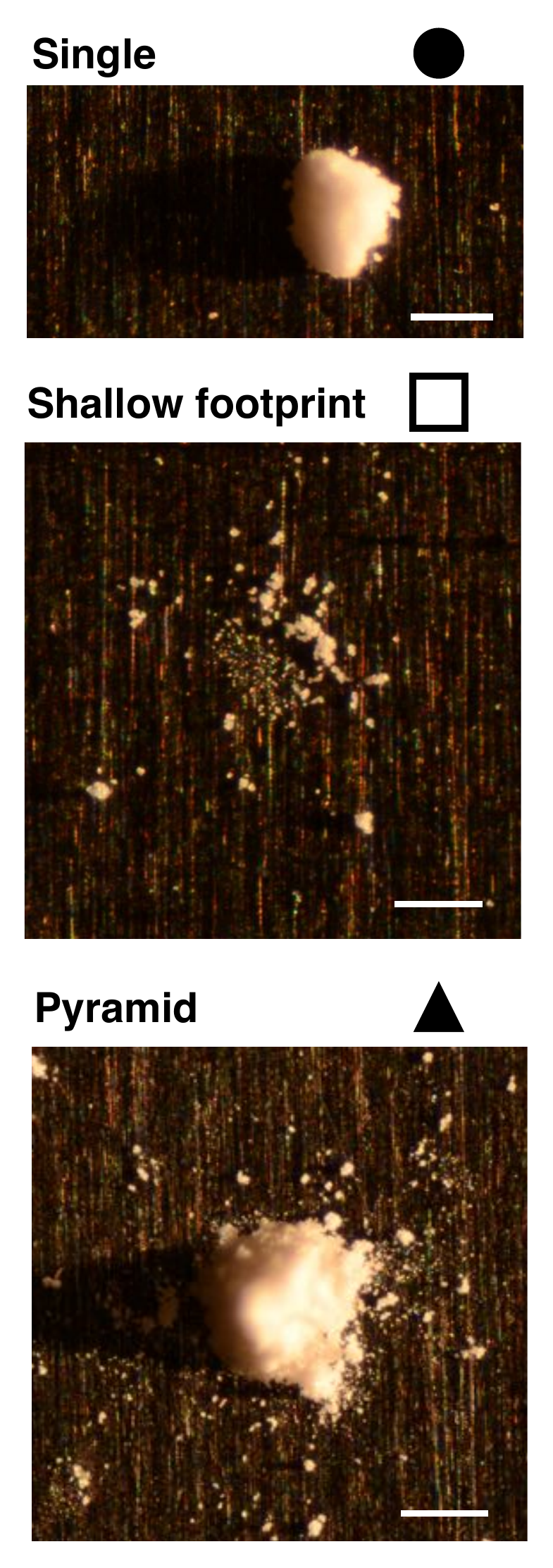}
	\includegraphics[height=0.8\columnwidth]{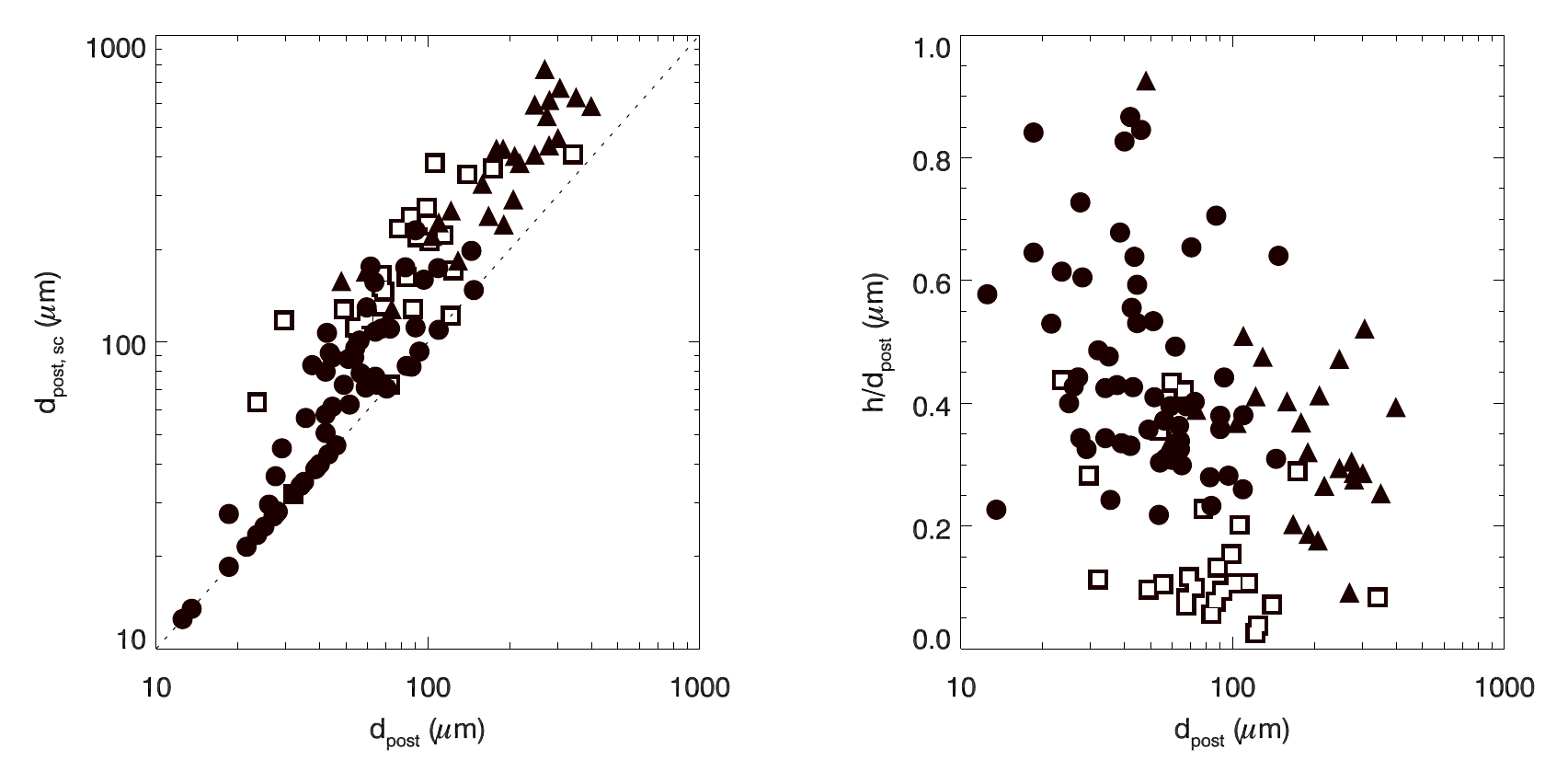}
    \caption{Deposit characteristics. \textit{Left:} Representative examples of the deposit morphology classes; the symbols indicated are the ones used in the diagrams in this and other figures. The scale bars are 100~\um~long. \textit{Middle:} Scatterfield diameter related to the deposit diameter; error bars are smaller than symbol sizes. \textit{Right:} Height-to-base ratio of the central fragment of the deposit. Symbols are plotted without error bars to improve readability; see Sect.~\ref{sec:methods:analysis} for a discussion on the uncertainties.}
    	\label{fig:morph}
	\end{figure*}
%, see Sec.~\ref{sec:analysis:postimpact}

The pre-impact and post-impact dimension measurements yield a rough estimate of the mass transfer fraction, defined as 
\begin{equation}
\mathrm{TF} = \frac{m_{\rm post}}{m_{\rm pre}}.
\label{eq:tf}
\end{equation}
As described above, $m_{\rm post}$ is systematically underestimated and $m_{\rm pre}$  is systematically overestimated (see Sect.~\ref{sec:analysis:preimpact}), especially for very small particles ($d_{\rm pre} \lesssim 50$~\um). For these particles the derived value for $\mathrm{TF}$ should be considered as a lower limit (also see Sect.\ref{sec:results:matrix}). For larger particles, propagation of the uncertainties in the measured properties leads to an error in the value of $\mathrm{TF}$ from 70 per cent for 70~\um-sized particles down to 15 per cent for 400~\um-sized particles. Hence, values of $\mathrm{TF}$ for small particles should only be considered as indicative. 

%\todo{in the end, we discard the values of TF for particles both small ($<50$~\um) and fast ($>4$~\ms).} 

Aside from these quantitative measurements, we also classified the qualitative morphology of the deposits. These classes are described in the next section. 

%\clearpage

% S9	S1	
% S11	S2
% S13  	S3
% S20 	S4
% S17	S5
% S15	S6
% S16 	S7

\section{Results}
\label{sec:results}

We present the results from experiments carried out on 7 targets, two of which were fired on twice (see Tab.~\ref{tab:experiments}). The experiments span a velocity range of 0.3~--~6~m~s$^{-1}$, which overlaps with the range measured on particles from 67P that entered the GIADA instrument \citep{Fulle2016, DellaCorte2016}. A total of 110 dust particles in the size range 30-410~$\mu$m could be traced on the video images and their deposits subsequently identified on the target. 

In Sect.~\ref{sec:results:morphologies}, we present the results of the post-impact analysis, introducing a categorization of the morphology of deposits. In Sect.~\ref{sec:results:matrix} we relate these post-impact results to the pre-impact properties of the dust particles (size and velocity). In Sect.~\ref{sec:results:collisions}, we present some typical collision types that we see leading to these morphologies. 

\subsection{Morphologies of deposits}
\label{sec:results:morphologies}

A variety of deposit morphologies was observed, with a range in scatter field diameter and height-to-base-ratio (Fig.~\ref{fig:morph}). Fig.~\ref{fig:morph} (\textit{right}) displays representative examples of three different morphological types that were observed. We defined a criterion to classify the deposits into three groups: `single', `shallow footprint' and `pyramid'. These are defined as follows:
\begin{itemize}
\item[-] \textit{Single} - Deposits with a clearly defined central component with few or no scattered fragments ($d_{\rm post, sc}/d_{\rm post} \lesssim 2$). Most single deposits found are small particles ($d\lesssim80$~\um), with a broad range in height-to-base ratio ($h/d_{\rm post} \sim 0.1-0.9$, more than 30 per cent exceeding 0.5).  As we will show in Sect.~\ref{sec:results:collisions}, many single particles are the result of sticking of the entire aggregate upon impact with little or no fragmentation.
\item[-] \textit{Shallow footprint} - A collection of monomers with no dominant central fragment ($d_{\rm post, sc} \gtrsim d_{\rm post}$). For these deposits the `central fragment' is defined by eye as the central area where the monomers are most tightly packed. The scattering field is on average about twice the size of the central fragment. The height-to-base ratio is low ($<0.2$ for most particles). Shallow footprints are the result of bouncing upon impact (see Sect.~\ref{sec:results:collisions}).
\item[-] \textit{Pyramid} - A clearly defined central fragment with scattered fragments around it ($d_{\rm post, sc} > d_{\rm post}$). The central fragment is generally pyramid-shaped, but in some cases has a flat or cratered upper part. The diameter of the scattering field is on average about twice the central deposit diameter. The central fragment is typically three times as wide as it is high. Pyramids are the result of fragmentation upon impact (see Sect.~\ref{sec:results:collisions}).
\end{itemize}
Note that these morphology classes are similar, but not identical, to those defined in \citet{Langevin2016} and \citet{Hornung2016}. We base our definition looking only at the results of our experiment; we do not assume \textit{a priori} that the deposit morphologies are the same as those detected by COSIMA.

% FIGURE: MATRIX
	\begin{figure}
	\includegraphics[width=0.87\columnwidth]{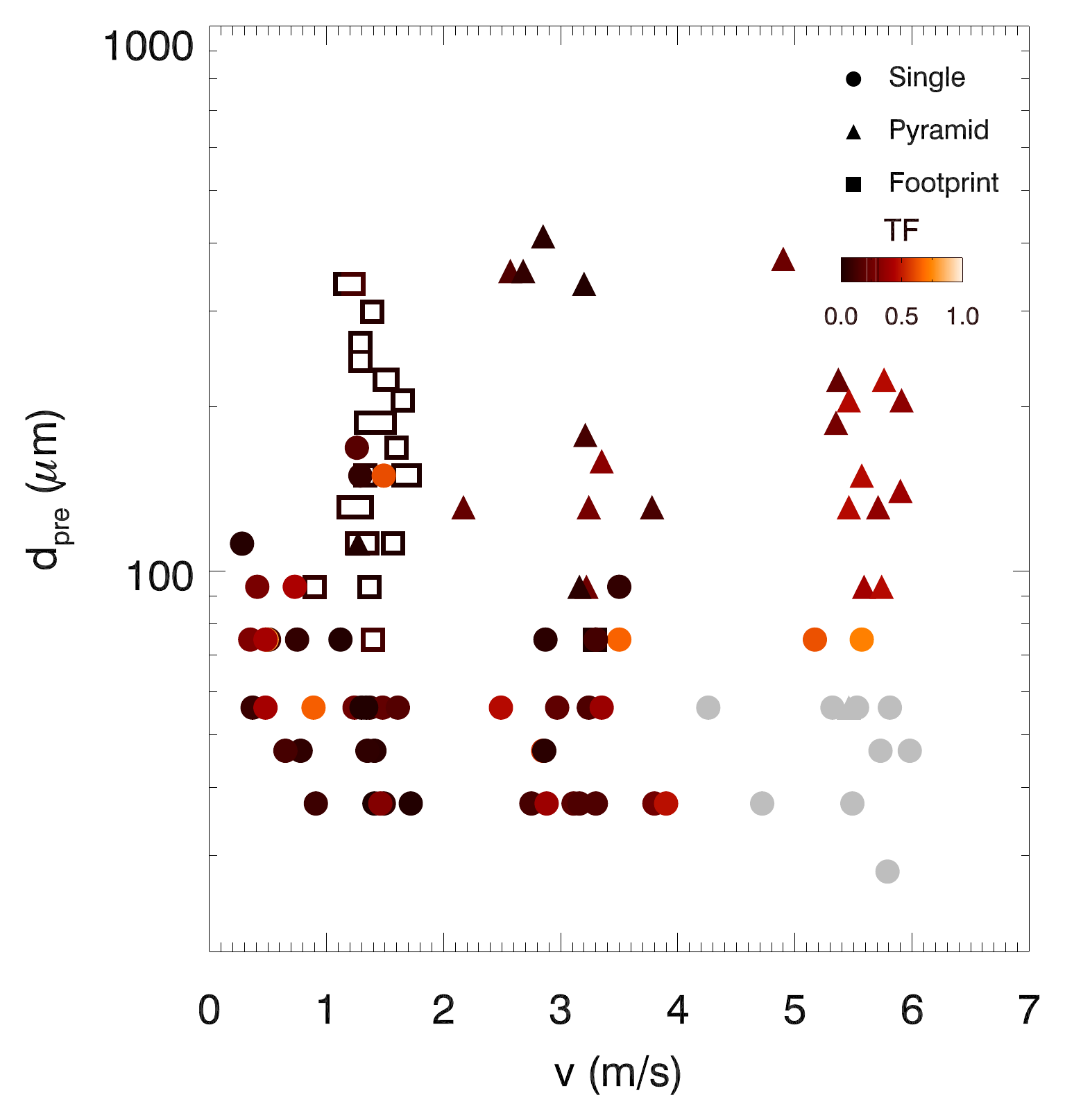}
%	\hspace*{10mm}
\begin{center}
\hspace*{-5mm}
	\includegraphics[width=0.75\columnwidth, page=2]{fig_collage_r_all.pdf}
\hspace*{-5mm}
	\includegraphics[width=0.75\columnwidth]{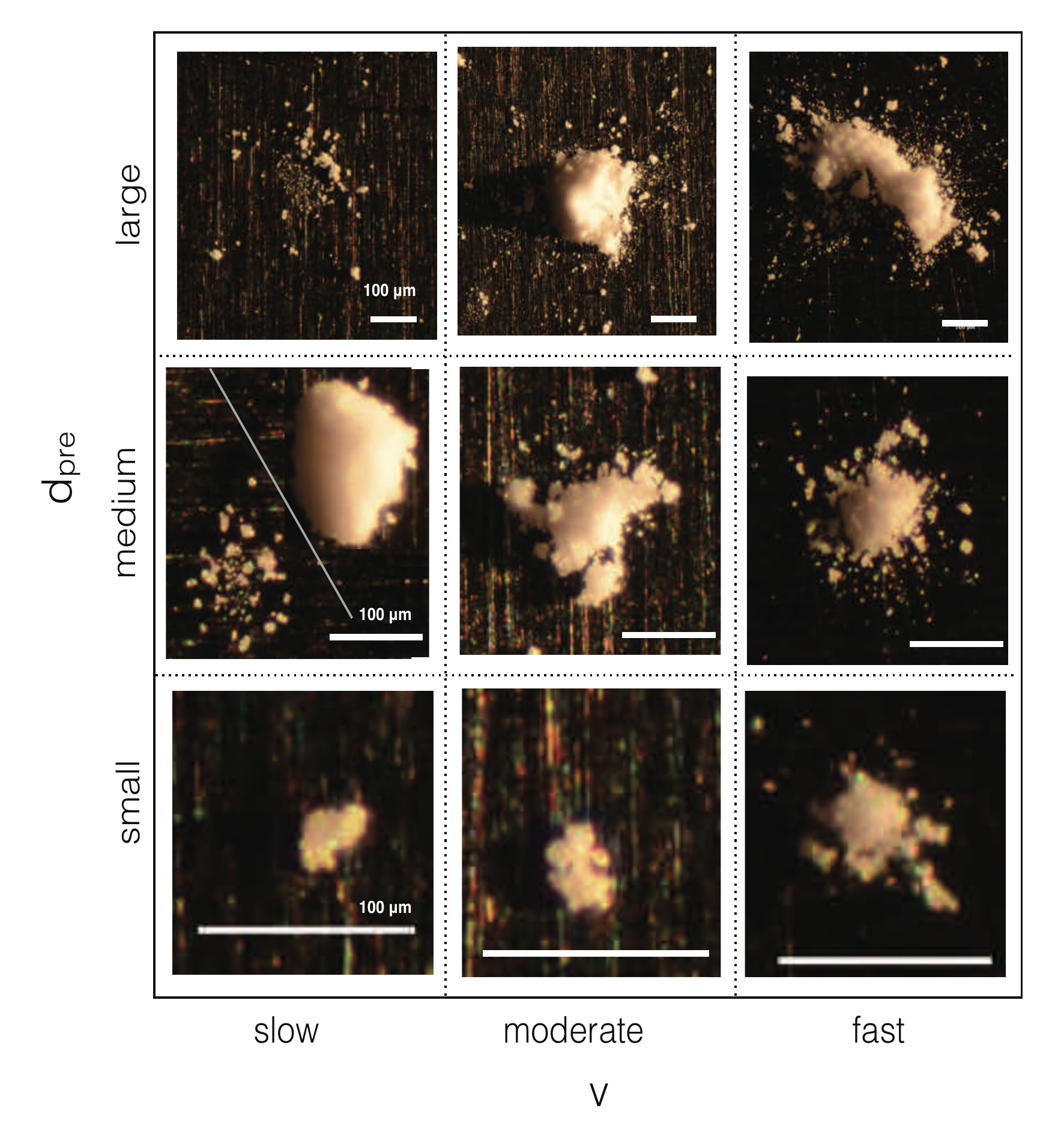}
\end{center}

    \caption{\textit{Top:} Morphologies and mass transfer function for particles used in the experiment, depending on particle size and velocity. Symbols are plotted without error bars to improve readability; see Sect.~\ref{sec:methods:analysis} for a discussion on the uncertainties. For particles represented by grey symbols, no meaningful value of $\mathrm{TF}$ could be derived (see text).
    \textit{Middle:} Summary of top panel, with sections indicating the type of collisions observed.
       \textit{Bottom:} Examples of some morphologies on same grid in ($d_{\rm pre},v$)-space. }
    	\label{fig:matrix}
	\end{figure}

\subsection{Relation of morphologies with particle size and velocity}
\label{sec:results:matrix}

The experimental results are summarised in Fig.~\ref{fig:matrix}. The top panel of this figure displays the parameter space in pre-impact particle size and velocity, with mass transfer function $\mathrm{TF}$ and morphology coded with colours and symbols. The middle panel is a schematic summary of the same diagram, with red sections indicating the collision types seen. The bottom panel shows microscopic images of representative examples of these types of deposits. 

A key insight provided by these results is that in most cases, and throughout the measured velocity range, only part of the particle sticks to the target. Only a few cases are seen with a value of $\mathrm{TF}$ that exceeds 0.6. Small particles ($d_{\rm pre}< 80~\mu$m) stick and leave `single' morphologies. For some of these deposits (represented with grey symbols in Fig.~\ref{fig:matrix}), no meaningful value of $\mathrm{TF}$ could be derived because of the limited resolution of the video images and the error introduced by the deconvolution method (as discussed in Sect.~\ref{sec:analysis:postimpact}). For most `single' deposits, the absence of scattered fragments suggests that $\mathrm{TF}$ is actually close to 1 (i.e. the particle stuck entirely). 

For larger particles ($80 < d_{\rm pre} < 410$~\um), two different cases are observed, depending on the impact velocity. At low velocities ($v<2$~m~s$^{-1}$) the majority of these particles leave deposits with a `shallow footprint' morphology ($\mathrm{TF}<0.2$); around ten per cent leave a `single' morphology. At velocities above 2~\ms, a `pyramid'-shaped deposit is left on the target ($0.2<\mathrm{TF}<0.5$). 

\begin{figure}
	\includegraphics[width=0.85\columnwidth]{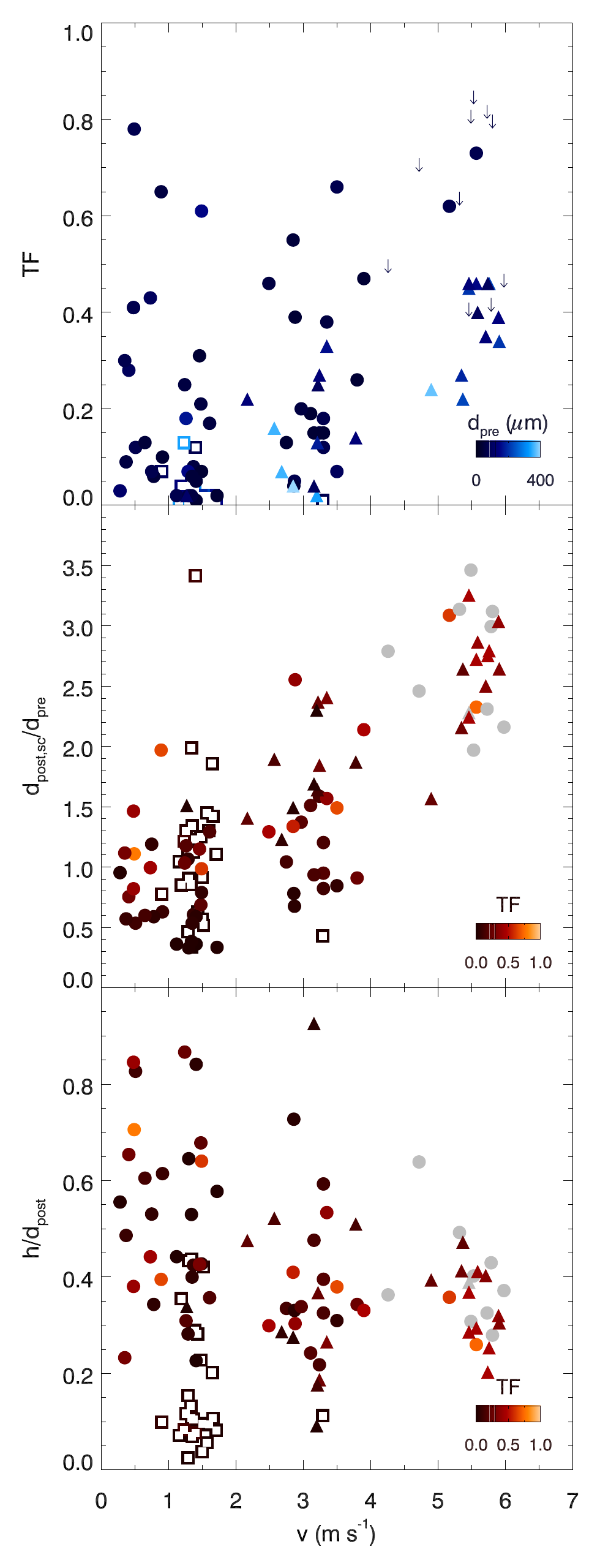}
    \caption{Properties of deposits. \textit{Top to bottom:} mass transfer function $\mathrm{TF}$ (colour traces size), relative scattering field size, and particle height-to-base ratio (colour traces $\mathrm{TF}$) as a function of impact velocity. Symbols are plotted without error bars to improve readability; see Sect.~\ref{sec:methods:analysis} for a discussion on the uncertainties. For particles represented with grey symbols no meaningful estimate could be made for $\mathrm{TF}$. The arrows in the top panel indicate upper limits.}
    	\label{fig:particleplots}
	\end{figure}

In Fig.~\ref{fig:particleplots}, some quantitative characteristics of the deposits are plotted as a function of impact velocity. It is shown that for pyramid-shaped deposits, $\mathrm{TF}$ and the degree of `spraying' (i.e. $d_{\rm post, sc}/d_{\rm pre}$) increase with $v$, regardless of particle size. Also, the fraction of the scatter field covered with particles increases with higher velocity, as larger fragments are scattered from the central deposit. 

 %FIGURE: PARTICLE STATS
%\clearpage

%FIGURE: COLLISIONS
	\begin{figure*}
	\includegraphics[width=1.9\columnwidth]{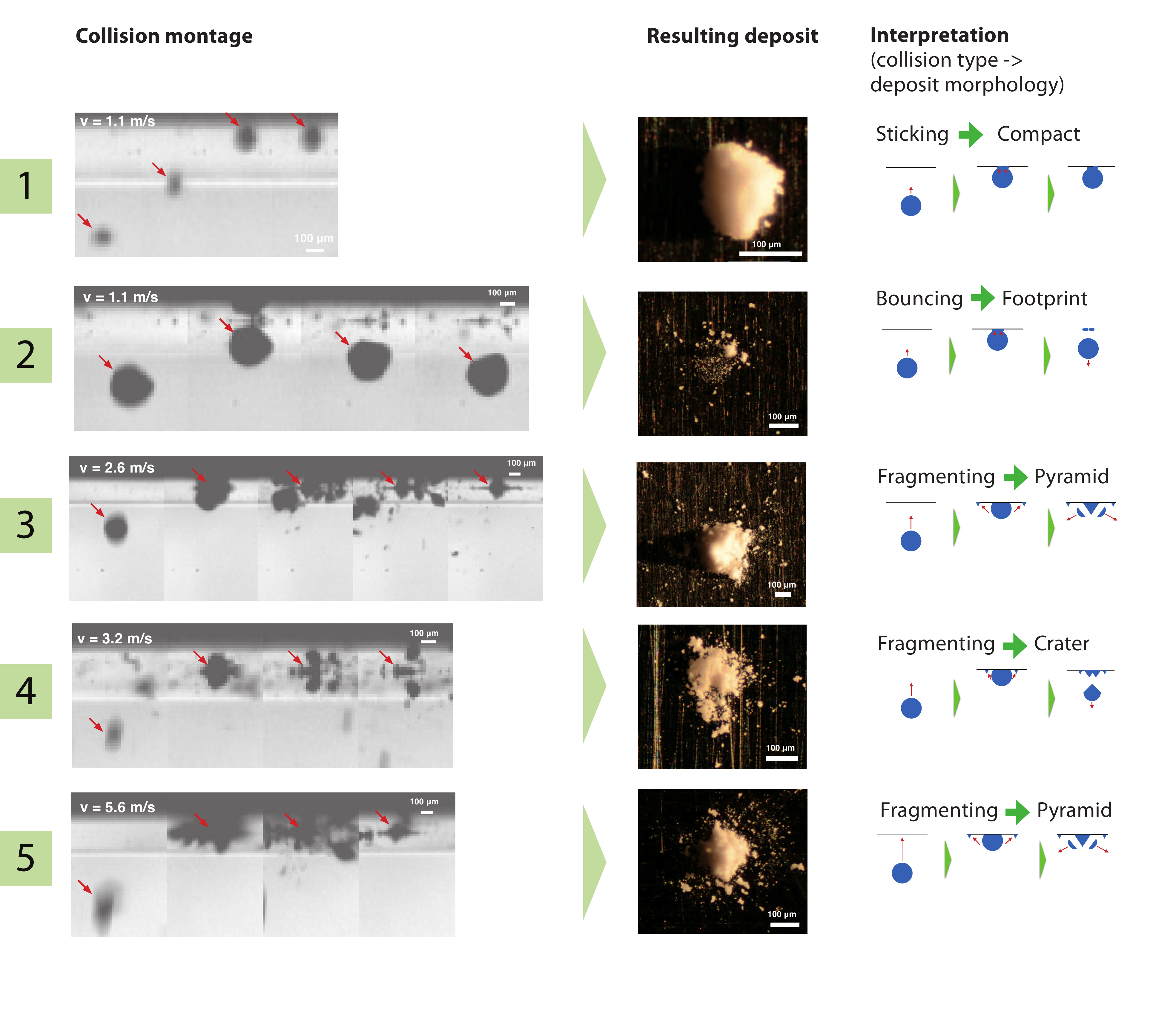}
    \caption{Different cases of collisions, and the deposits they leave on the target. From left to right, the columns display a montage of video images covering the collision, the resulting deposit on the target, and a schematic interpretation of the collision dynamics (red arrows indicate velocity vectors of particles and fragments). From top to bottom, the pre-impact velocity increases.}
    	\label{fig:collisions}
	\end{figure*}
	
We observe a difference in deposit height-to-base ratio ($h/d_{\rm post}$) between slow and fast particles. At low velocities ($v<2$~\ms), small particles that leave single morphologies have a wide range in height-to-base ratio, ($0.1<h/d_{\rm post}<0.9$). Slow particles larger than $\sim80$~\um~leave shallow footprints with a height-to-base ratio lower than 0.2. For deposits of fast particles ($v>2$~\ms), the values of the height-to-base ratio are seen to converge towards values in the range $0.2<h/d_{\rm post}<0.5$. Above $v=5.0$~\ms, no `outliers' are seen outside this range. This shows that even at high velocities, a particle is not compressed completely flat onto the surface. The lack of height-to-base ratios exceeding 0.5 for single deposits produced at high velocities is a consequence of the central component being damaged during the collisions, rather than flattening and/or compression; also see next subsection. 

\subsection{Collision types leading to different morphologies}
\label{sec:results:collisions}

In the previous subsections we have described the outcome of dust collisions on target surfaces, using tracing and size measurement of the dust particles \textit{before} impact. However, the video images also capture several frames during the individual collisions, allowing us to explain the observed morphology classes by collision physics. 

Fig.~\ref{fig:collisions} displays five different examples of dust particle collisions, representative of the entire parameter space in size and velocity. We see three different collision types occurring: sticking, bouncing and fragmentation.  In a sticking collision (case 1), the entire particle is seen to stick to the target, with few or no fragments ejected from the particle. Sticking is observed at low velocities and with small particles ($d_{\rm pre}<80$~\um), and leads to deposits of the `single' morphological type. Sticking collisions are only seen to occur at velocities $<2$~\ms.

A bouncing collision (case 2) leaves a shallow footprint at the location where the particle made contact with the target surface. During such collisions, some small fragments are ejected from the sides and front of the particle and create a scatter field on the target, but the bulk of the particle stays intact after it rebounds. Put in a more quantitative way, the parameter $\mu$, defined as the mass of the largest fragment divided by the mass of the original particle \citep{Guttler2010}, is close to 1. After the collision, the large fragment moves in the opposite direction at a velocity of typically half the impact velocity (i.e. coefficient of restitution $\sim0.5$). The deposit left on the target is seen to be a shallow footprint if the pre-impact particle diameter exceeds $d_{\rm pre} = 80$~\um. Below this value particles leave single deposits. Like sticking collisions, bouncing collisions are only seen to occur at velocities $<2$~\ms. 

Cases 3--5 display representative examples (with increasing impact velocity) of fragmenting collisions, which occur at velocities exceeding 2~\ms and leave a deposit with the `pyramid' morphology. In most cases, the particle breaks up into multiple small fragments ($\mu<0.2$), which are ejected from the target in all radial directions from the impact location; see the montage in Fig.~\ref{fig:preimpact}, bottom, for an example. During the collision, when the front of the particle makes contact with the target, the particle breaks up. Some momentum is transferred to the back of the particle, diverting the trajectory of the fragments. This results in a diverging motion of fragments, leading to a `scatter field' somewhat larger than the original particle. The scatter field is seen to be larger at higher impact velocities. The central deposit left after a fragmenting collision is in most cases pyramid-shaped and with a low height-to-base ratio (0.2-0.5). In some cases (e.g., case 4), a large fragment breaks off the central deposit, leaving a depression at its centre. Some small ($d_{\rm pre} < 80$~\um) particles at velocities $> 2$~\ms~leave deposits classified as `single', as they do not have a significant scatter field. These particles have a height-to-base ratio lower than 0.5, with the central component consisting of multiple parts which are still connected. Both of these properties indicate that a (simple) fragmenting collision likely took place. 

% \todo{Perhaps the collision type can be included as a parameter in the analysis. Would be a lot of work to do, and not every collision was isolated enough to monitor, but interesting nonetheless. Alternative: do a fragmentation analysis on these case examples. Downside: will still take up a lot of time, and lead to conclusions already found by other authors.}

%	\begin{figure*}
%	\includegraphics[width=\textwidth]{fig_matrix_collage.pdf}
%    \caption{Same as Fig.~\ref{fig:matrix}, with some examples of morphologies.}
%    	\label{fig:matrix_collage}
%	\end{figure*}

\section{Discussion}
\label{sec:discussion}

\subsection{Comparison with other collision experiments}
\label{sec:discussion:physics}

The results of our experiments compare well to previous studies of dust aggregate collisions \citep{BlumWurm2008, Guttler2010}. The experiments in these reviews mainly study particle-particle collisions and growth mechanisms, while our experiments focus on the deposits left on static targets. 

Our experiment compares best to Experiment 18 in \citet{Guttler2010}, where SiO$_2$ aggregates were accelerated in a vacuum tube and collided upon a glass plate. The parameter space in aggregate size and velocity encompasses the ranges used in our experiment. The material used, however, is different: \citet{Guttler2010} use amorphous aggregates built of monodispersed monomers of diameter $\sim1.5$~\um, with a lower material density ($\rho_{\rm m}\sim2 \times 10^3$\kgm) and volume filling factor ($\phi=0.15$). The tensile strength is $\sim10^3$~Pa, similar to the material in our experiment. Despite the differences in material density and structure, the experiments summarised in \citet{Guttler2010} observe a similar barrier between sticking/bouncing and fragmentation ($v\sim2$~\ms~for small aggregates). This indicates that within this parameter range, the value of the fragmentation barrier is determined by the value of the tensile strength, despite the difference in monomer size distribution, density and volume filling factor, and consistent with numerical predictions of the fragmentation barrier by \citet{DominikTielens1997}. The critical fragmentation velocity of a few metres per second is thus a robust result that does not depend strongly on the shape of the monomers, nor on their size with the range 0.1-10~\um. However, for aggregates of monodisperse monomers below this value, the surface energy may be enhanced leading to a higher fragmentation barrier. This barrier may be increased further if the particles are coated by organic material \citep{BlumWurm2000}; this may have consequences for the comparison of our experiments with cometary dust (see Sect.~\ref{sec:discussion:material}).

With respect to sticking and/or bouncing, \citet{Guttler2010} find an enhanced sticking probability at velocities of 0.3~--~5~\ms~for aggregates of a mass of $10^{-8}-10^{-6}$~g. This range in mass is equivalent to particles with $d_ {\rm pre} < 50$~\um~in our experiments, corrected for the difference in $\rho_{\rm b}$. Indeed, the smallest particles in our experiments are also shown to have a higher sticking rate. 

Compaction is expected to be negligible upon a single sticking event at velocities of order 1~\ms \citep[][who use monodispersed material with $\phi \sim 0.15$]{Weidling2009}. As the dynamical pressure increases with the square of the impact velocity, above the fragmentation barrier compaction is expected to be larger with $\phi$ increasing with a factor up to 1.5 upon a single collision \citep{Blum2006}. Thus, the single particles created after a low-velocity sticking collision are not compacted, but the pyramid-shaped deposits may be. As the effect of compaction increases for material with lower volume filling factors \citep{Teiser2011, Meru2013}, cometary material may be more compactable than our laboratory sample. 

%In the \citet{Guttler2010} experiment, the fragmentation probability of a grain of mass $m$ is seen to decrease from $m=5\times10^{-7}$~g at $v=0.8$~\ms to $m=1\times10^{-7}$~g at $v=7$~\ms. 

\subsection{Comparison between sample material and cometary dust}
\label{sec:discussion:material}

The test material used in the experiments presented in this study differs from cometary dust in several ways. Some of these differences are expected to affect the sticking properties of the material, and hence the outcome of the collection onto the target plates. In this subsection, we compare the density, structure and composition of our test material with the known properties of cometary dust, and discuss their effect on the outcome of the collection. 

The bulk density of the test material $\rho_{\rm b}$ = ($0.9 \pm 0.1$) $\times 10^{3}$~\kgm~is in the same order of magnitude as the bulk density of the `compact' dust particles in the coma of 67P, as estimated based on GIADA measurements \citep[$\rho_{\rm b} \sim 1-2 \times 10^3$~\kgm,][]{Rotundi2015, Fulle2016}. The `fluffy' dust particles have a derived bulk density that is 3 orders of magnitude lower than this \citep[$\rho_{\rm b}\sim1$~\kgm,][]{Fulle2015}. 

The strength and sticking properties are affected by the size distribution and the volume filling factor of the material. The dust aggregates are tightly packed, while cometary dust may have a more fluffy structure. The strength of the test material is of the same order of the strength of 67P dust, which is estimated at $\sim 10^3$~Pa \citep{Hornung2016}. However, the test particles are randomly packed and thus have a nonhierarchical inner structure. This may result in different collision and sticking properties than cometary dust, e.g. the `porous aggregates' described by \citet{Mannel2016}. Also, the shape and roughness of the particle can be critical for stickiness and tensile strength \citep[see e.g., ][]{Poppe2000a, Castellanos2005} 
%Castellanos: contact area (and thus adhesion) scales as ratio r_asperity/r_particle

The structure and resulting volume filling factor of the cometary dust is a parameter not well constrained directly from observations. While GIADA provides a bulk density of individual particles inferred from cross section and momentum measurements, the material density and volume filling factor remain degenerate and can only be constrained by indirect methods (see e.g., \citealt{Schulz2015, Fulle2015}). \citet{Hornung2016} estimate the volume filling factor by measurements of the fragment sizes of the `rubble pile'-like deposits, and assumptions on the packing of their parent particles before entering the instrument funnel. These authors arrive at values in the range $\phi \sim 0.4-0.6$ for the volume filling factor.  Comparing measurements of composition from the SIMS mass spectrometer with the observed volume of the deposits may provide further constraints on the material density, and hence volume filling factor, of the coma dust population. 

Compared to the dust in the coma, the comet nucleus has a lower bulk density estimated at $\sim 0.5 \times 10^3$~\kgm, and a lower volume filling factor $\sim$ 0.15-0.30, based on measurements by the CONSERT and RSI experiments \citep{Kofman2015, Patzold2016}. This lower value may be explained by a higher macroporosity and higher abundance of ices in the nucleus. The composition, structure and volume filling factor of the test sample is an interesting parameter to vary in future experiments, to establish how this value correlates with the outcome of collisions on detector surfaces. 

The presence of ice in the cometary dust particles is expected to increase its stickiness by one order of magnitude \citep{Gundlach2011, Gundlach2015, Blum2014, Aumatell2014}. However, the dust particles collected by COSIMA at altitudes of 10-100 km \citep{Langevin2016}, are thought to have shed their volatiles and consist solely of organic and mineral refractory components. This is indicated by the absence of obvious signs of ice having evaporated from particles on the target surface \citep{Schulz2015}. Therefore, while the ice content of the dust may strongly affect the entry of dust particles into the coma, the absence of ice in our experiments is representative of the circumstances during dust collection in the spacecraft. 

%\todo{can you spend one sentence to say why there is no ice anymore?}
%\todo{MSB: Literature: what fraction of the dust ejected contains ice? If there are few secondary sources (?) observed, then probably little for 67P?}
%Guilbert-Lepoutre, A. et al. Pre-perihelion activity of comet 67P/Churyumov- Gerasimenko. Astron. Astrophys. 567, http://dx.doi.org/10.1051/0004-6361/ 201424186 (2014).
%A grain composed of an iceÐmineral mixture would not shatter at low-velocity collection; instead, the icy part of such a grain would evaporate very shortly after collection, leaving one or more voids in the particle that remains on the target plate. Grains composed of (nearly) pure water-ice would evaporate at or shortly after collection and create a dark signature on the target plate. At the scale of the COSIMA image resolution (pixel size is 14 mm), there is no hint of volatiles having left the grains after collection. In other words, there is no indication of an iceÐmineral mixture, or of pure icy grains hitting the target. This is in contrast to cometary grains remotely observed, or collected before the Rosetta mission.

The composition of cometary dust may also affect the sticking properties. The coma dust of 67P is rich in silicates, but, unlike our sample, also contains carbon and organics \citep{Hilchenbach2016, Fray2016}. Organics may enhance stickiness if the material softens by melting \citep[see e.g.,][]{BlumWurm2000}; however, this strongly depends on the type of organic material and the ambient temperature \citep[][Sect. 6.2]{Guttler2010}. Therefore, more knowledge of the composition of the 67P dust is needed to evaluate the influence of organics on the collision physics inside the Rosetta instruments.

%naa; be careful, organics have a broad range of melting temperatures.
%Just say that organics may stick better, but there are also a lot of organic materials which are not stickier, like charcola etc
%is sentence is in contradiction to your paragraphs before: ice is more sticky, organics maybe more sticky?

If the dust particles in our sample are charged, this may effectively lower the fragmentation barrier, as the electrical force between the electrons works to break apart the aggregate. For the particle sizes, velocities and short trajectories used in our experiments, this effect is expected to be negligible \citep{Poppe2000b}. In cometary dust, however, there may be a non-negligible charge of the dust particles captured by Rosetta, as suggested by the interpretation of \citet{Fulle2015}. This may affect the sticking probability of these aggregates on target surfaces, and possibly lower the fragmentation barrier.

Finally, the optical properties of our test material differ from those of cometary dust. The test material is white and thus has a much higher albedo than the 67P dust particles detected by COSIMA (albedo = 1-40 per cent, \citealt{Langevin2016}), which is caused by carbonaceous components of the latter. The difference in albedo should be taken into account when comparing the experimental results with COSIMA data. 

	%FIGURE: COSISCOPECOMPARE
	\begin{figure}
	\includegraphics[width=\columnwidth]{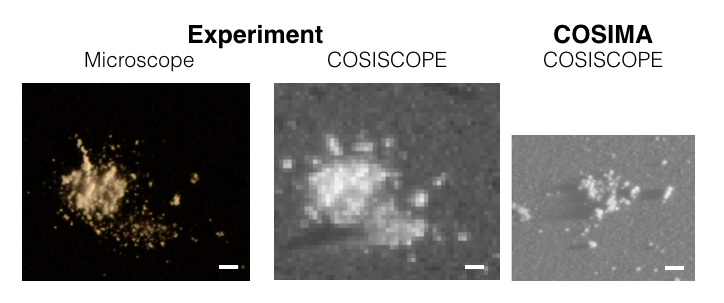}
    \caption{Particle deposit created in collision experiment ($v=2.6$~\ms) on gold black target, subsequently studied with an optical microscope (\textit{left}) and the flight spare of COSISCOPE (\textit{middle}). \textit{Right}: Cometary dust particle (`shattered cluster') detected by COSIMA \citep[adapted from][]{Merouane2016}. Scale bars indicate 100~\um.}
    	\label{fig:cosiscope_compare}
	\end{figure}

%FIGURE: BLACKGOLDMIDAS
	\begin{figure}
	\includegraphics[width=\columnwidth]{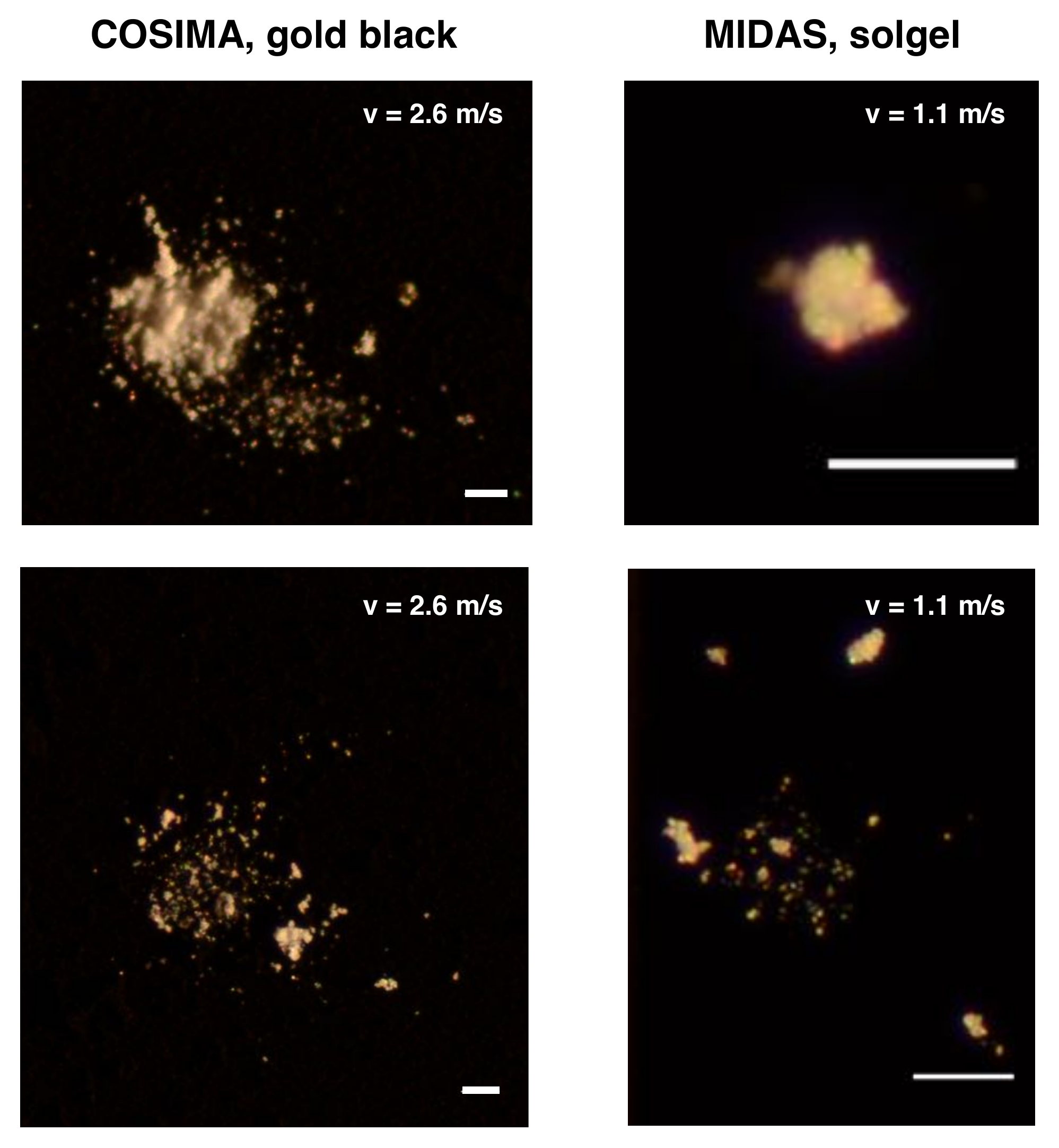}
    \caption{Two deposits of SiO$_2$ particles after collisions on COSIMA gold black targets (left, $v=2.6$~\ms) and MIDAS solgel targets (right, $v=1.1$~\ms). Scale bars indicate 100~\um.}
    	\label{fig:g1m1}
	\end{figure}

\subsection{Towards an interpretation of Rosetta data}
\label{sec:discussion:comparison}

% Qualitative comparison with Rosetta data
% Prospects for using this tool to make a quantitative interpretation of R data

The aim of our series of collision experiments is to better constrain the properties of the coma dust population from COSIMA and MIDAS data. It is tempting to directly compare our results to the Rosetta images. However, while the images may look the same on inspection, we know that some of the properties of cometary dust and experimental conditions that differ from the situation at the spacecraft, as described in the previous subsection. It is therefore premature to conclude that there is a one-to-one correspondence of our experimental results to COSIMA data. In this subsection, we summarise the qualitative results of our experiments applicable to Rosetta results, the implications that may be drawn from them, and suggestions for future experiments to confirm these hypotheses.

Most of the different morphological types that we detect qualitatively resemble those seen by COSIMA. Firstly, the `single' deposits in our experiments resemble the particles referred to as `compact' by \citet{Langevin2016} whose nomenclature refers to the optical morphology, and as `no breakup' or `simple breakup' by \citet{Hornung2016}, whose nomenclature refers to the collision type that has likely produced the particle. Some of the `single' deposits in our experiments consist of multiple fragments that are stuck together, but have little or no fragments scattered around them. These bear more resemblance to the `glued cluster' particles defined by \citep{Langevin2016}.

Secondly, the `pyramid' deposits in our experiments resemble the `rubble pile' and `scattered cluster' morphologies (referred to as `catastrophic breakup' by \citealt{Hornung2016}, whose nomenclature refers to the collision type). The third morphological type, the `shallow footprint', has no obvious counterpart in the COSIMA data. Only a few examples exist of COSIMA particles with a very low height-to-base ratio. These particles have a low contrast on the COSISCOPE image, which may be caused by a low albedo, but can also be because the pixel is filled by a collection of non-bound small particles that do not fill the pixel of the COSISCOPE imager. 

The aggregate particle detected by MIDAS (particle E, \citealt{Bentley2016}) bears resemblance to the `shallow footprints' in our experiments. However, the MIDAS particle has connections between subunits, while in some of our shallow footprints larger spaces exist between subunits. Furthermore, MIDAS can only detect parts of deposits with a height less than at maximum  10~\um~above the detector surface, usually less. It cannot be excluded that some partially imaged particles have greater heights than measured. This is unlikely to apply to particle E, however, as additional scans of the target show that the regions surrounding the partly imaged particle are rather empty (although the particle itself was removed from the target during the initial scan).

The main insight drawn from this initial run of experiments is the fact that typically, only part of the particle sticks to the target surface; the appearance of this deposit is determined by the impact velocity. Upon a bouncing or fragmenting collision, the fragments reverse their direction and do not end up on the target plate.  The bulk of the particle is lost in the detector. Future experiments may confirm whether this also applies to the dust collected in Rosetta, or that a difference in structure results in `collapse' of the particle as described by \citet{Hornung2016}. Interpreting all COSIMA particles as containing all the mass of the entire original grain may thus be misleading. It should be noted that the micron-sized, isolated particles detected by MIDAS \citep{Bentley2016} are in fact single, undamaged particles (or fragments of a larger particle that fragmented in the funnel), as sticking is to be expected at these small sizes. %Also, it may be interesting to check for shallow footprints in the Rosetta data.

Apart from the known and estimated differences between our test material and cometary dust, as described in Sect.~\ref{sec:discussion:material}, there are some apparent optical differences between our deposits and the MIDAS and COSIMA particles. The first difference is the size and shape of the individual monomers. The largest monomer in our dust aggregates is 10~\um, while the six MIDAS particles published so far do not show a constituent single component larger than 3.3~\um. The shape of the laboratory monomers is irregular (and thus random), while the MIDAS grains are elongated. These properties may affect the collision properties; they can be varied in future experiments. 

Our experiments seem to suggest that deposits detected by COSIMA (and MIDAS) are only part of the particle, except the `compact' / `undamaged' particles. However, as described above, there are known differences between our sample material and comet dust, and the Rosetta environment, whose impact have the potential to change this interpretation. Therefore, more detailed studies looking into the relevant parameters are necessary. In a next run of experiments, we will start varying some of these properties, to explore their effect on collisions, sticking properties and deposit morphology. Also, we will use different imaging techniques (e.g. using the replica of the on-board COSISCOPE instrument, see Fig.~\ref{fig:cosiscope_compare}) to facilitate a better comparison to Rosetta measurements. As monomer size is an insightful and the most feasible parameter to vary, this will be done in a next series of experiments.

Looking forward to these future experiments, we executed two test experiments on a COSIMA gold black and a MIDAS solgel target Fig.~\ref{fig:g1m1}. Qualitatively, all different morphology classes were retrieved. This suggests that the sticking properties of the gold black, solgel and silver target surfaces are comparable and that the choice of target material does not have a major effect on the collisional outcome. This hypothesis may be tested with future experiments, in order to make a direct comparison to MIDAS and COSIMA data.

\section{Summary and Conclusions}
\label{sec:conclusions}

We performed an initial run of experiments, in which we collided polydisperse SiO$_2$ aggregates in the size range 30~--~410~\um~with a target surface at impact velocities of 0.3~--~6~\ms. The experiments aim to simulate the circumstances of dust collection in the COSIMA and MIDAS instruments on the Rosetta spacecraft. Our main conclusions are:

\begin{itemize}
\item[-] At $v<2$~\ms~and $d_{\rm pre}<80$~\um, sticking of complete particles leads to a `single' deposit. 
\item[-] At $v<2$~\ms~and $d_{\rm pre}>80$~\um, a bouncing collision leads to a `shallow footprint' deposit. 
\item[-] At $v>2$~\ms~and $d_{\rm pre}>80$~\um, a fragmenting collision leads to pyramid-shaped deposit with fragments scattered around it, and a loss of part of the initial dust particle. The amount of mass transferred to the target increases with the impact velocity. 
\item[-] The `single' and pyramid-shaped deposits are very reminiscent of the morphological types observed by COSIMA; a closer examination is needed to allow a quantitative comparison.
\item[-] Despite the different deposit morphologies, the projectiles in the experiment were similar in shape and composition; only velocities were varied. This implies that velocity is the main driver of the appearance of the deposits on the target. 
\item[-] It is likely that bouncing and fragmenting collisions have occurred in the Rosetta instruments, resulting in sticking of only part of the particle that entered the instrument. This should be taken into account when analysing the morphology and quantity of cometary dust particles. 
\item[-] The differences in composition and structure of the test sample and cometary dust may affect both the collision physics and the appearance of deposits. Future experiments will explore the parameter space, providing trends on how these properties (density, packing, monomer size, composition) and the resulting specific surface energy affect the morphology of deposits. This will allow us to establish to what degree these conclusions are applicable to the Rosetta results.
\end{itemize}

\section*{Acknowledgements}

This work has been financially supported by NWO, project no. ALW-GO/15-01. MSB and TM acknowledge funding by the Austrian Science Fund FWF P 28100-N36. The authors acknowledge the support of the technical staff at TU Braunschweig in building the experimental setup. Ren\'{e} Weidling, Rainer Schr\"{a}pler, Horst Uwe Keller and Rens Waters are acknowledged for useful discussions. The SEM image (Fig.~\ref{fig:sample}, bottom) was taken at the Electron Microscopy Center Amsterdam, Academic Medical Center, Amsterdam, The Netherlands. 
 
%[backup microscope teams at UvA and AMC?]

%%%%%%%%%%%%%%%%%%%%%%%%%%%%%%%%%%%%%%%%%%%%%%%%%%



%\bibliographystyle{mnras}

\begin{thebibliography}{}
\makeatletter
\relax
\def\mn@urlcharsother{\let\do\@makeother \do\$\do\&\do\#\do\^\do\_\do\%\do\~}
\def\mn@doi{\begingroup\mn@urlcharsother \@ifnextchar [ {\mn@doi@}
  {\mn@doi@[]}}
\def\mn@doi@[#1]#2{\def\@tempa{#1}\ifx\@tempa\@empty \href
  {http://dx.doi.org/#2} {doi:#2}\else \href {http://dx.doi.org/#2} {#1}\fi
  \endgroup}
\def\mn@eprint#1#2{\mn@eprint@#1:#2::\@nil}
\def\mn@eprint@arXiv#1{\href {http://arxiv.org/abs/#1} {{\tt arXiv:#1}}}
\def\mn@eprint@dblp#1{\href {http://dblp.uni-trier.de/rec/bibtex/#1.xml}
  {dblp:#1}}
\def\mn@eprint@#1:#2:#3:#4\@nil{\def\@tempa {#1}\def\@tempb {#2}\def\@tempc
  {#3}\ifx \@tempc \@empty \let \@tempc \@tempb \let \@tempb \@tempa \fi \ifx
  \@tempb \@empty \def\@tempb {arXiv}\fi \@ifundefined
  {mn@eprint@\@tempb}{\@tempb:\@tempc}{\expandafter \expandafter \csname
  mn@eprint@\@tempb\endcsname \expandafter{\@tempc}}}

\bibitem[\protect\citeauthoryear{{Aumatell} \& {Wurm}}{{Aumatell} \&
  {Wurm}}{2014}]{Aumatell2014}
{Aumatell} G.,  {Wurm} G.,  2014, \mn@doi [\mnras] {10.1093/mnras/stt1921},
  \href {http://adsabs.harvard.edu/abs/2014MNRAS.437..690A} {437, 690}

\bibitem[\protect\citeauthoryear{{Bentley} et~al.,}{{Bentley}
  et~al.}{2016}]{Bentley2016}
{Bentley} M.~S.,  et~al., 2016, \mn@doi [\nat] {10.1038/nature19091}, \href
  {http://adsabs.harvard.edu/abs/2016Natur.537...73B} {537, 73}

\bibitem[\protect\citeauthoryear{{Blum} \& {Schr{\"a}pler}}{{Blum} \&
  {Schr{\"a}pler}}{2004}]{Blum2004}
{Blum} J.,  {Schr{\"a}pler} R.,  2004, \mn@doi [Physical Review Letters]
  {10.1103/PhysRevLett.93.115503}, \href
  {http://adsabs.harvard.edu/abs/2004PhRvL..93k5503B} {93, 115503}

\bibitem[\protect\citeauthoryear{{Blum} \& {Wurm}}{{Blum} \&
  {Wurm}}{2000}]{BlumWurm2000}
{Blum} J.,  {Wurm} G.,  2000, \mn@doi [\icarus] {10.1006/icar.1999.6234}, \href
  {http://adsabs.harvard.edu/abs/2000Icar..143..138B} {143, 138}

\bibitem[\protect\citeauthoryear{{Blum} \& {Wurm}}{{Blum} \&
  {Wurm}}{2008}]{BlumWurm2008}
{Blum} J.,  {Wurm} G.,  2008, \mn@doi [\araa]
  {10.1146/annurev.astro.46.060407.145152}, \href
  {http://adsabs.harvard.edu/abs/2008ARA%26A..46...21B} {46, 21}

\bibitem[\protect\citeauthoryear{{Blum}, {Schr{\"a}pler}, {Davidsson}  \&
  {Trigo-Rodr{\'{\i}}guez}}{{Blum} et~al.}{2006}]{Blum2006}
{Blum} J.,  {Schr{\"a}pler} R.,  {Davidsson} B.~J.~R.,
  {Trigo-Rodr{\'{\i}}guez} J.~M.,  2006, \mn@doi [\apj] {10.1086/508017}, \href
  {http://adsabs.harvard.edu/abs/2006ApJ...652.1768B} {652, 1768}

\bibitem[\protect\citeauthoryear{{Blum}, {Gundlach}, {M{\"u}hle}  \&
  {Trigo-Rodriguez}}{{Blum} et~al.}{2014}]{Blum2014}
{Blum} J.,  {Gundlach} B.,  {M{\"u}hle} S.,   {Trigo-Rodriguez} J.~M.,  2014,
  \mn@doi [\icarus] {10.1016/j.icarus.2014.03.016}, \href
  {http://adsabs.harvard.edu/abs/2014Icar..235..156B} {235, 156}

\bibitem[\protect\citeauthoryear{{Castellanos}}{{Castellanos}}{2005}]{Castellanos2005}
{Castellanos} A.,  2005, \mn@doi [Advances in Physics]
  {10.1080/17461390500402657}, \href
  {http://adsabs.harvard.edu/abs/2005AdPhy..54..263C} {54, 263}

\bibitem[\protect\citeauthoryear{{Colangeli} et~al.,}{{Colangeli}
  et~al.}{2007}]{Colangeli2007}
{Colangeli} L.,  et~al., 2007, \mn@doi [\ssr] {10.1007/s11214-006-9038-5},
  \href {http://adsabs.harvard.edu/abs/2007SSRv..128..803C} {128, 803}

\bibitem[\protect\citeauthoryear{{Della Corte} et~al.,}{{Della Corte}
  et~al.}{2015}]{DellaCorte2015}
{Della Corte} V.,  et~al., 2015, \mn@doi [\aap] {10.1051/0004-6361/201526208},
  \href {http://adsabs.harvard.edu/abs/2015A%26A...583A..13D} {583, A13}

\bibitem[\protect\citeauthoryear{{Della Corte} et~al.,}{{Della Corte}
  et~al.}{2016}]{DellaCorte2016}
{Della Corte} V.,  et~al., 2016, \mn@doi [\mnras] {10.1093/mnras/stw2529},
  \href {http://adsabs.harvard.edu/abs/2016MNRAS.462S.210D} {462, S210}

\bibitem[\protect\citeauthoryear{{Dominik} \& {Tielens}}{{Dominik} \&
  {Tielens}}{1997}]{DominikTielens1997}
{Dominik} C.,  {Tielens} A.~G.~G.~M.,  1997, \mn@doi [\apj] {10.1086/303996},
  \href {http://adsabs.harvard.edu/abs/1997ApJ...480..647D} {480, 647}

\bibitem[\protect\citeauthoryear{{Dominik}, {Blum}, {Cuzzi}  \&
  {Wurm}}{{Dominik} et~al.}{2007}]{Dominik2007}
{Dominik} C.,  {Blum} J.,  {Cuzzi} J.~N.,   {Wurm} G.,  2007, Protostars and
  Planets V, \href {http://adsabs.harvard.edu/abs/2007prpl.conf..783D} {pp
  783--800}

\bibitem[\protect\citeauthoryear{{Fray} et~al.,}{{Fray}
  et~al.}{2016}]{Fray2016}
{Fray} N.,  et~al., 2016, \mn@doi [\nat] {10.1038/nature19320}, \href
  {http://adsabs.harvard.edu/abs/2016Natur.538...72F} {538, 72}

\bibitem[\protect\citeauthoryear{{Fulle} et~al.,}{{Fulle}
  et~al.}{2015}]{Fulle2015}
{Fulle} M.,  et~al., 2015, \mn@doi [\apjl] {10.1088/2041-8205/802/1/L12}, \href
  {http://adsabs.harvard.edu/abs/2015ApJ...802L..12F} {802, L12}

\bibitem[\protect\citeauthoryear{{Fulle} et~al.,}{{Fulle}
  et~al.}{2016}]{Fulle2016}
{Fulle} M.,  et~al., 2016, \mn@doi [\mnras] {10.1093/mnras/stw2299}, \href
  {http://adsabs.harvard.edu/abs/2016MNRAS.462S.132F} {462, S132}

\bibitem[\protect\citeauthoryear{{Gundlach} \& {Blum}}{{Gundlach} \&
  {Blum}}{2015}]{Gundlach2015}
{Gundlach} B.,  {Blum} J.,  2015, \mn@doi [\apj] {10.1088/0004-637X/798/1/34},
  \href {http://adsabs.harvard.edu/abs/2015ApJ...798...34G} {798, 34}

\bibitem[\protect\citeauthoryear{{Gundlach}, {Kilias}, {Beitz}  \&
  {Blum}}{{Gundlach} et~al.}{2011}]{Gundlach2011}
{Gundlach} B.,  {Kilias} S.,  {Beitz} E.,   {Blum} J.,  2011, \mn@doi [\icarus]
  {10.1016/j.icarus.2011.05.005}, \href
  {http://adsabs.harvard.edu/abs/2011Icar..214..717G} {214, 717}

\bibitem[\protect\citeauthoryear{{G{\"u}ttler}, {Krause}, {Geretshauser},
  {Speith}  \& {Blum}}{{G{\"u}ttler} et~al.}{2009}]{Guttler2009}
{G{\"u}ttler} C.,  {Krause} M.,  {Geretshauser} R.~J.,  {Speith} R.,   {Blum}
  J.,  2009, \mn@doi [\apj] {10.1088/0004-637X/701/1/130}, \href
  {http://adsabs.harvard.edu/abs/2009ApJ...701..130G} {701, 130}

\bibitem[\protect\citeauthoryear{{G{\"u}ttler}, {Blum}, {Zsom}, {Ormel}  \&
  {Dullemond}}{{G{\"u}ttler} et~al.}{2010}]{Guttler2010}
{G{\"u}ttler} C.,  {Blum} J.,  {Zsom} A.,  {Ormel} C.~W.,   {Dullemond} C.~P.,
  2010, \mn@doi [\aap] {10.1051/0004-6361/200912852}, \href
  {http://adsabs.harvard.edu/abs/2010A%26A...513A..56G} {513, A56}

\bibitem[\protect\citeauthoryear{{Hilchenbach} et~al.,}{{Hilchenbach}
  et~al.}{2016}]{Hilchenbach2016}
{Hilchenbach} M.,  et~al., 2016, \mn@doi [\apjl] {10.3847/2041-8205/816/2/L32},
  \href {http://adsabs.harvard.edu/abs/2016ApJ...816L..32H} {816, L32}

\bibitem[\protect\citeauthoryear{{Hornung} et~al.,}{{Hornung}
  et~al.}{2014}]{Hornung2014}
{Hornung} K.,  et~al., 2014, \mn@doi [\planss] {10.1016/j.pss.2014.08.011},
  \href {http://adsabs.harvard.edu/abs/2014P%26SS..103..309H} {103, 309}

\bibitem[\protect\citeauthoryear{{Hornung} et~al.,}{{Hornung}
  et~al.}{2016}]{Hornung2016}
{Hornung} K.,  et~al., 2016, \mn@doi [\planss] {10.1016/j.pss.2016.07.003},
  \href {http://adsabs.harvard.edu/abs/2016P%26SS..133...63H} {133, 63}

\bibitem[\protect\citeauthoryear{{Johansen}, {Blum}, {Tanaka}, {Ormel},
  {Bizzarro}  \& {Rickman}}{{Johansen} et~al.}{2014}]{Johansen2014}
{Johansen} A.,  {Blum} J.,  {Tanaka} H.,  {Ormel} C.,  {Bizzarro} M.,
  {Rickman} H.,  2014, \mn@doi [Protostars and Planets VI]
  {10.2458/azu_uapress_9780816531240-ch024}, \href
  {http://adsabs.harvard.edu/abs/2014prpl.conf..547J} {pp 547--570}

\bibitem[\protect\citeauthoryear{{Kissel} et~al.,}{{Kissel}
  et~al.}{2007}]{Kissel2007}
{Kissel} J.,  et~al., 2007, \mn@doi [\ssr] {10.1007/s11214-006-9083-0}, \href
  {http://adsabs.harvard.edu/abs/2007SSRv..128..823K} {128, 823}

\bibitem[\protect\citeauthoryear{{Kofman} et~al.,}{{Kofman}
  et~al.}{2015}]{Kofman2015}
{Kofman} W.,  et~al., 2015, \mn@doi [Science] {10.1126/science.aab0639}, \href
  {http://adsabs.harvard.edu/abs/2015Sci...349b0639K} {349}

\bibitem[\protect\citeauthoryear{{Kothe}, {Blum}, {Weidling}  \&
  {G{\"u}ttler}}{{Kothe} et~al.}{2013}]{Kothe2013}
{Kothe} S.,  {Blum} J.,  {Weidling} R.,   {G{\"u}ttler} C.,  2013, \mn@doi
  [\icarus] {10.1016/j.icarus.2013.02.034}, \href
  {http://adsabs.harvard.edu/abs/2013Icar..225...75K} {225, 75}

\bibitem[\protect\citeauthoryear{{Langevin} et~al.,}{{Langevin}
  et~al.}{2016}]{Langevin2016}
{Langevin} Y.,  et~al., 2016, \mn@doi [\icarus] {10.1016/j.icarus.2016.01.027},
  \href {http://adsabs.harvard.edu/abs/2016Icar..271...76L} {271, 76}

\bibitem[\protect\citeauthoryear{{Lorek}, {Gundlach}, {Lacerda}  \&
  {Blum}}{{Lorek} et~al.}{2016}]{Lorek2016}
{Lorek} S.,  {Gundlach} B.,  {Lacerda} P.,   {Blum} J.,  2016, \mn@doi [\aap]
  {10.1051/0004-6361/201526565}, \href
  {http://adsabs.harvard.edu/abs/2016A%26A...587A.128L} {587, A128}

\bibitem[\protect\citeauthoryear{{Mannel}, {Bentley}, {Schmied}, {Jeszenszky},
  {Levasseur-Regourd}, {Romstedt}  \& {Torkar}}{{Mannel}
  et~al.}{2016}]{Mannel2016}
{Mannel} T.,  {Bentley} M.~S.,  {Schmied} R.,  {Jeszenszky} H.,
  {Levasseur-Regourd} A.~C.,  {Romstedt} J.,   {Torkar} K.,  2016, \mn@doi
  [\mnras] {10.1093/mnras/stw2898}, \href
  {http://adsabs.harvard.edu/abs/2016MNRAS.462S.304M} {462, S304}

\bibitem[\protect\citeauthoryear{{Meisner}, {Wurm}  \& {Teiser}}{{Meisner}
  et~al.}{2012}]{Meisner2012}
{Meisner} T.,  {Wurm} G.,   {Teiser} J.,  2012, \mn@doi [\aap]
  {10.1051/0004-6361/201219099}, \href
  {http://adsabs.harvard.edu/abs/2012A%26A...544A.138M} {544, A138}

\bibitem[\protect\citeauthoryear{{Merouane} et~al.,}{{Merouane}
  et~al.}{2016}]{Merouane2016}
{Merouane} S.,  et~al., 2016, \mn@doi [\aap] {10.1051/0004-6361/201527958},
  \href {http://adsabs.harvard.edu/abs/2016A%26A...596A..87M} {596, A87}

\bibitem[\protect\citeauthoryear{{Meru}, {Geretshauser}, {Sch{\"a}fer},
  {Speith}  \& {Kley}}{{Meru} et~al.}{2013}]{Meru2013}
{Meru} F.,  {Geretshauser} R.~J.,  {Sch{\"a}fer} C.,  {Speith} R.,   {Kley} W.,
   2013, \mn@doi [\mnras] {10.1093/mnras/stt1447}, \href
  {http://adsabs.harvard.edu/abs/2013MNRAS.435.2371M} {435, 2371}

\bibitem[\protect\citeauthoryear{{P{\"a}tzold} et~al.,}{{P{\"a}tzold}
  et~al.}{2016}]{Patzold2016}
{P{\"a}tzold} M.,  et~al., 2016, \mn@doi [\nat] {10.1038/nature16535}, \href
  {http://adsabs.harvard.edu/abs/2016Natur.530...63P} {530, 63}

\bibitem[\protect\citeauthoryear{{Poppe}, {Blum}  \& {Henning}}{{Poppe}
  et~al.}{2000a}]{Poppe2000a}
{Poppe} T.,  {Blum} J.,   {Henning} T.,  2000a, \mn@doi [\apj]
  {10.1086/308626}, \href {http://adsabs.harvard.edu/abs/2000ApJ...533..454P}
  {533, 454}

\bibitem[\protect\citeauthoryear{{Poppe}, {Blum}  \& {Henning}}{{Poppe}
  et~al.}{2000b}]{Poppe2000b}
{Poppe} T.,  {Blum} J.,   {Henning} T.,  2000b, \mn@doi [\apj]
  {10.1086/308631}, \href {http://adsabs.harvard.edu/abs/2000ApJ...533..472P}
  {533, 472}

\bibitem[\protect\citeauthoryear{{Riedler} et~al.,}{{Riedler}
  et~al.}{2007}]{Riedler2007}
{Riedler} W.,  et~al., 2007, \mn@doi [\ssr] {10.1007/s11214-006-9040-y}, \href
  {http://adsabs.harvard.edu/abs/2007SSRv..128..869R} {128, 869}

\bibitem[\protect\citeauthoryear{{Rotundi} et~al.,}{{Rotundi}
  et~al.}{2015}]{Rotundi2015}
{Rotundi} A.,  et~al., 2015, \mn@doi [Science] {10.1126/science.aaa3905}, \href
  {http://adsabs.harvard.edu/abs/2015Sci...347a3905R} {347, aaa3905}

\bibitem[\protect\citeauthoryear{{Schulz} et~al.,}{{Schulz}
  et~al.}{2015}]{Schulz2015}
{Schulz} R.,  et~al., 2015, \mn@doi [\nat] {10.1038/nature14159}, \href
  {http://adsabs.harvard.edu/abs/2015Natur.518..216S} {518, 216}

\bibitem[\protect\citeauthoryear{{Teiser}, {Engelhardt}  \& {Wurm}}{{Teiser}
  et~al.}{2011}]{Teiser2011}
{Teiser} J.,  {Engelhardt} I.,   {Wurm} G.,  2011, \mn@doi [\apj]
  {10.1088/0004-637X/742/1/5}, \href
  {http://adsabs.harvard.edu/abs/2011ApJ...742....5T} {742, 5}

\bibitem[\protect\citeauthoryear{{Weidling}, {G{\"u}ttler}, {Blum}  \&
  {Brauer}}{{Weidling} et~al.}{2009}]{Weidling2009}
{Weidling} R.,  {G{\"u}ttler} C.,  {Blum} J.,   {Brauer} F.,  2009, \mn@doi
  [\apj] {10.1088/0004-637X/696/2/2036}, \href
  {http://adsabs.harvard.edu/abs/2009ApJ...696.2036W} {696, 2036}

\bibitem[\protect\citeauthoryear{{Weidling}, {G{\"u}ttler}  \&
  {Blum}}{{Weidling} et~al.}{2012}]{Weidling2012}
{Weidling} R.,  {G{\"u}ttler} C.,   {Blum} J.,  2012, \mn@doi [\icarus]
  {10.1016/j.icarus.2011.10.002}, \href
  {http://adsabs.harvard.edu/abs/2012Icar..218..688W} {218, 688}

\makeatother
\end{thebibliography}


\bibliographystyle{mnras}

\input{footprint_arxiv.bbl}
\bsp	% typesetting comment
\label{lastpage}
\end{document}